\documentclass[12pt]{report}
\usepackage{amsfonts}
\usepackage{amsmath}
\usepackage{import}
\usepackage{amssymb}
\usepackage[pdftex]{graphicx}					
\usepackage[font=small,margin=50pt]{caption}
\usepackage{xspace}
\usepackage{float}
\usepackage{subfigure}
\usepackage[authoryear]{natbib}
\usepackage{color}
\usepackage[margin=2.5cm]{geometry}	
\usepackage{graphicx}
\usepackage{wrapfig}
\usepackage{enumitem}
\usepackage{array}
\usepackage{dcolumn}
\usepackage{booktabs}
\usepackage{listings}
\usepackage{setspace}
\usepackage{nicefrac}
\usepackage{siunitx}
\newcommand{\msun}{M_{\odot}}

\newcommand{\ms}{\text{ ms}^{-1}}
\newcommand{\unit}[1]{\mathbf{\hat{#1}}}

\makeatletter
\newcommand\ackname{Acknowledgements}
\if@titlepage
	\newenvironment{acknowledgements}{%
    	\titlepage
        \null\vfil
        \@beginparpenalty\@lowpenalty
        \begin{center}%
        	\bfseries \ackname
            \@endparpenalty\@M
        \end{center}}%
        {\par\vfil\null\endtitlepage}
\else
	\newenvironment{acknowledgements}{%
    	\if@twocolumn
        	\section*{\abstractname}%
        \else
            \small
            \begin{center}%
            	{\bfseries \ackname\vspace{-.5em}\vspace{\z@}}%
            \end{center}%
            \quotation
        \fi}
        {\if@twocolumn\else\endquotation\fi}
\fi
\makeatother

\begin{document}
\pagenumbering{roman}
	\thispagestyle{empty}

\newgeometry{margin=3.5cm}

\begin{titlepage}
\center

\vfill
\Huge {A Compact Spectrograph to Search for Extrasolar Planets}
\vfill
\large  {Carlos H. Bacigalupo} \\

\vfill
\vfill
October, 2012
\vfill
\vfill
\large A thesis submitted to Macquarie University in partial completion of the requirements of the degree of Bachelor of Advanced Science with Honours
\vfill

\includegraphics{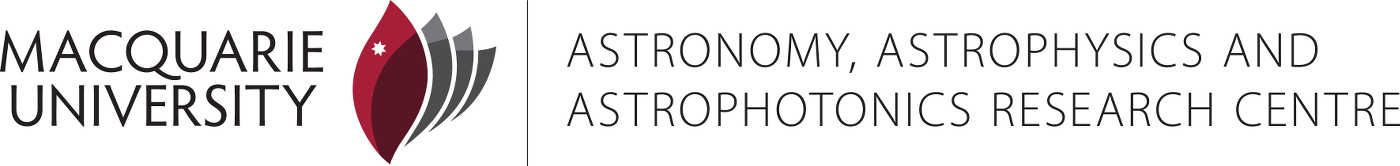}

\end{titlepage}

\newgeometry{margin=2.5cm}
	\clearpage
\begin{center}
To Susan and Hannah
\end{center}
\thispagestyle{empty}
\pagebreak

\begin{abstract}
The most successful method used so far to search for extrasolar planets is the radial velocity technique, where periodical shifts on the measured emission from a star provide evidence for an orbiting planet. This method has been used on large telescopes with large and expensive instrumentation, only enabling a small amount of observing time per star. We have developed a compact spectrograph fed by one or several single-mode fibres that avoids the need for complex fibre scrambling or gas absorption cells for calibration. In principle, this will enable planet searches around bright stars over the next few years. We aim to pave the way for large networks of small telescopes searching for Earth-like planets. At a resolving power of R$\sim$50000, I have characterized this spectrograph, determined its stability and the fidelity required for a simultaneous calibration source. 
\end{abstract}

\begin{acknowledgements}
I want to thank my supervisor, Dr. Michael Ireland, who has led me through the mysterious paths of the optical world. 

For their support during a very special year, to my friends from the honours group: Andrew, Vincent, Chris, Jacob, James and Shane. Suffering is a lot easier in a group. 

To the new members of the Macquarie family, Tobias and Joao, for joining the project at the right moment to help me reach important milestones.

Last, and certainly not least, to my daughter Hannah who every day expands my universe at accelerating rates and to my wife Susan whose endless love and coffee got me through this honours year.
\end{acknowledgements}

\pagebreak

\thispagestyle{empty}
	\tableofcontents{}
	\thispagestyle{empty}
	\clearpage
	\pagenumbering{arabic}	
	\chapter{Introduction}

The discovery of the first accepted planet outside our solar system by~\cite{wolszczan_planetary_1992} soon to be followed by~\citep{mayor_jupiter-mass_1995}, marked the beginning of a new era in astronomy. Our preconceived notions of planetary systems have been consistently challenged by new discoveries since then. We are now familiar with extreme configurations such as giant planets at surprisingly short distances from their host stars~\citep{mayor_jupiter-mass_1995}, multi-planetary systems around binary stars~\citep{orosz_kepler-47:_2012} and a claimed Earth-mass planet around our closest star system~\citep{dumusque_earth-mass_2012}. As technology advances, increasing the precision of the instrumentation used for detection, new and more varied configurations have been discovered leading to a wider understanding of planetary systems. However, despite the improvements and efforts invested in the field, little is known about the evolution of these systems and how they relate to the properties of their host star. 

\begin{figure}[ht]
	\centering
		\includegraphics{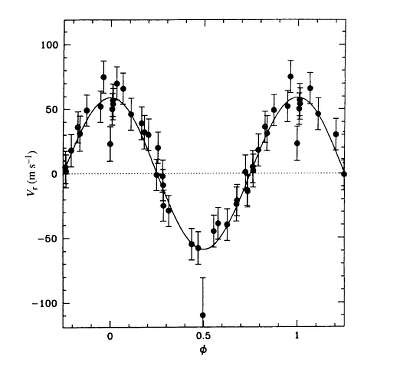}
		\caption{Orbital motion of 51 Peg, the first sun-like star with an identified planet-sized companion~\citep{mayor_jupiter-mass_1995}. The vertical axis represents the star's velocity as seen from Earth and the horizontal axis shows the orbital period.}
		\label{fig:RV}
\end{figure}

From the several methods used to detect planets around other stars, or exoplanets, two have produced the majority of the results. The transit method looks for periodical dips in the luminosity of the host star as the indication of the presence of a planet and the radial velocity method (RV) measures the to and fro motion of the host star as it is pulled by the gravitational attraction of the orbiting the planet. The latter, also known as Doppler spectroscopy, is responsible for the discovery of more than 500 planets around solar-type stars.

The purpose of this project is to build and characterise an optical instrument to achieve the level of precision required for the detection of exoplanets by the radial velocity method using off-the-shelf components. This approach can pave the way for a replicable model that would allow simultaneous observations from different points on Earth. This low-cost solution can be appealing to institutions around the world leading to a network of small telescopes feeding a single centralised data centre. 

The instrument that allows us to achieve such a level of precision is a spectrograph, an optical arrangement that disperses the light from a star into the different colours that constitute it. The collection of colours or wavelengths, known as the spectrum, is projected on a Charged Coupled Device\footnote{Charged Coupled Devices or CCDs can convert photons into electrons to be read and stored as digital values.} to be studied. Key information on the composition of the star can be learned by analysing the variation of intensities across its spectrum. A moving star will shift the intensity pattern as it moves through the line of sight, the line that connects the object with the observer. This is a consequence of the stretching or contracting of the emitted waves as the source moves, it is known as the Doppler effect. With careful analysis, the change in the spectrum over time can reveal the presence of a companion, see Figure~\ref{fig:RV}. The Replicated High-resolution Exoplanet and Asteroseismology~(RHEA) spectrograph is the instrument we developed for this purpose.

There are limiting factors that determine our capacity to detect exoplanets. These limitations are related to both the technology available for the construction of our instruments as well as the properties of exoplanetary systems themselves. Stellar atmospheres are dynamic environments that cloak the signature of exoplanets. This ``stellar noise'' plays a significant role in determining the characteristics of our observing methods and data analysis. 

Stars whose outer layers are dominated by convective currents can excite solar-like oscillations. The star acts as a resonant box and, analogous to a musical instrument, its physical characteristics determine the period and amplitude of the dominant oscillations. In sun-like stars, oscillations have typical periods of a few minutes and amplitudes of a fraction of a meter per second. The period of these oscillations scales with the square root of the mean density of the star and the radial velocity amplitude scales with the luminosity-to-mass ratio~\citep{christensen-dalsgaard_physics_2004}. These relations indicate that the characteristics of the oscillations will evolve with the star. 

Stars that reach the end of their hydrogen burning phase expand to become red giants. When the mass of this class of stars ranges between one and two solar masses, and the radius is close to  ten times the radius of the Sun, they produce oscillations of similar amplitude than the radial velocity expected from the presence of planets. This creates challenges in separating the signals from the stars' intrinsic noise to the gravitational pull created by the planet. Similarly, the changes in radial velocity generated by the influence of a planet depend on its distance to the host star. This difference can allow us to discriminate between intrinsic stellar oscillations and the signature of an exoplanet. To effectively separate both signals,   observations need to be performed for long enough periods to average out the intrinsic stellar noise. In the case of red giants, this means that each observing session can last from a few hours to a full night. This requirement makes it impractical for a large telescope to become available for the required observational time, as a large number projects compete for such resources. 

The precision in the RV method has been largely improved by the use of simultaneous known reference sources~\citep{butler_attaining_1996}. The use of lamps containing well identified  spectral lines allows us to project a sample beam of known characteristics though our spectrograph. The advantage of a known source is the identified lines, or brightest wavelengths, and their relative intensities. When projected though the spectrograph, they can help us identify the precise location on the detector of key reference points that can be used as guides to later read the stars' spectrum. 
The reference source can be placed in the path of the light coming from the telescope, effectively inserting the reference lines in the spectrum observed. The is commonly achieved by the use of iodine cells. The advantage is a simultaneous image containing both spectra and eliminating the need to compensate for changes occurred between each acquisition. The clear disadvantage is the loss introduced by the addition of an element in the light path. Alternatively, the known reference can be sampled independently and a wavelength scale model developed to read the location of the individual wavelengths on the CCD. This is the approach adopted for this project. The sources used are mercury (Hg) and thorium argon (ThAr), although the latter has weak reference lines and can be very faint requiring long exposure times to be detected. 

\subsubsection{Science}

The constant development in this dynamic field presents plenty of opportunities. Some of the key questions that this honours project aims to help investigate in the subsequent stages are the relation between stellar mass and the properties of the planetary systems; do planets survive the red giant phase? How do they affect the host star during this process?
\cite{soker_rotation_2000} suggest that the range of properties observed in red giants is a consequence of the transfer of angular momentum from engulfed planets. Later stages of this project will be capable of furthering the observations that can support or refute that statement. 
Can a statistical survey of the local neighbourhood agree with the predictions of galactic archaeology projects like~\cite{freeman_new_2002}?

\subsubsection{Outline of Thesis}

This thesis describes the instrumental and software development necessary to achieve the precision required for the detection of exoplanets.

Chapter~2 is a review of the evolution of exoplanetary search; the techniques used and the results achieved at the moment of publication.

An analysis of the practicalities of exoplanetary search is the focus of Chapter~3. The limitations on the instrumentation and the availability of resources is balanced to build a case for the use of small telescopes feeding compact spectrographs.  

Chapter~4 is a description of the instrumentation developed for this project, an overview of the different components of the optical system. This chapter is to become the first version of a reference manual paving the way for a replicable model.

Chapter~5 describes the software components. The Wavelength Scale Model (WSM) is the module that computes the optical path that a monochromatic beam produces as it travels through the system; it is the key component of the calibration package. The fitting module is an optimization add-on that iterates over the WSM to find the optimal input parameters. Thermal and pressure changes can be detected by the sub-pixel shift module which also extends the precision to the level required to detect exoplanets. 

In Chapter~6 a quantitative analysis and derivation of the results obtained is presented.

The different possibilities that arise from the results of this project are analysed in Chapter~7 with particular emphasis on the steps required to take the developed prototype into a replicable precision astronomical instrument.
	\chapter{Exoplanets}
The possibility of planets outside our solar system has generated interest and stimulated our imagination throughout human history. Conjectures on the chances of our planetary system being just one amongst many others can be traced back to figures such as Democritus, Giordano Bruno and Sir Isaac Newton. In the year 300 BCE, Epicurus suggested that ``There are infinite worlds both like and unlike this world of ours. For the atoms being infinite in number... are borne on far out into space.''~\citep{epicurus_epicurus_1926}. \cite{huygens_celestial_1698} stated that ``the Earth may justly liken'd to the planets...[which have] Men...[that] chiefly differ from Beasts in the study of Nature...[and who] have Astronomy''. It was only in the last 20 years that speculations of this nature became scientific facts~\citep{wolszczan_planetary_1992}. 

The first scientific claim of a planet beyond the solar system belongs to~\cite{jacob_certain_1855} around the binary star system 70 Ophiuchi. The failure to compute an orbit that matched the observations led to the proposal of a planet. This suggestion was later discredited by~\cite{moulton_limits_1899}. This was the first of several failed attempts to identify an exoplanet. In the 1940s, a series of giant planet claims proved to be the consequence of  false signals. \cite{van_de_kamp_astrometric_1963} announced a Jupiter mass planet orbiting Barnard's star on a 24 year period. This claim turned out to be a systematic error introduced by the instrument. 

Despite the unsuccessful experimental results, theoretical work was slowly building the models that would eventually guide the research. \cite{struve_proposal_1952} presented the possibility of a Jupiter-like planet as close as 2\% of the distance between the Earth and the Sun, the Astronomical Unit (AU)\footnote{1 Astronomical Unit=1.496 $\times$ 10$^{11}$ m}. Although the technology required for such measurements was just reaching maturity at the time, no observations to confirm or deny such claims were attempted. We now know that such a configuration is not only possible but common. Our anthropocentric views, still fixed in the patterns suggested by our own solar system, arguably delayed the first confirmed observation by decades. 

After several unconfirmed attempts, it was in an unexpected location that the first exoplanets were found. In 1992, the discovery of a two planet system orbiting a neutron star arguably became the first confirmed observations of a planet outside our solar system~\citep{wolszczan_planetary_1992}. Neutron stars have consumed all of their fuel and have collapsed due to their own gravity. They are held by neutron degeneracy pressure\footnote{Particles in degeneracy pressure are subject to the Pauli exclusion principle where only one particle of a given type can occupy the same quantum state.} preventing them from collapsing further. These can be fast rotating stars as the conservation of angular momentum dictates that the frequency of rotation increases as the radius decreases. When the star has a strong magnetic field, the process of contraction increases the density of the field lines. These two effects lead to a fast rotating star with a very localized emission known as a pulsar. It was around a star of this kind that the first exoplanets were discovered. The star PSR B1257+12 in the constellation of Virgo is a millisecond pulsar and became the first known host star of an exoplanetary system. It is now known to have 3 planets orbiting it. This early discovery presented the question of planet survivability around highly evolved stars. One of the goals of the RHEA spectrograph is to address key questions of this kind. Only 3 years later the first planet around a Sun-like star by~\cite{mayor_jupiter-mass_1995} was found.

These milestones formally inaugurated an era of exploration currently totalling 809 confirmed exoplanets and more than 2300 candidates~\citep{batalha_planetary_2012}.

The current search is mainly focused on Earth-like planets orbiting the habitable zone\footnote{Also known as the `Goldilocks Zone', as an analogy to the children's story 'Goldilocks and the 3 Bears' due to its ideal conditions for life.}, the narrow region that holds the right conditions for a terrestrial-like planet to hold liquid water on its surface~\citep{kasting_habitable_1993}. The challenges to overcome are several and despite a rapid growth, detection technologies are still in an early stage. Results are biased by the instrumental limitations and generalized conclusions are difficult to reach. Nonetheless, this is a golden era in the search for exoplanets and certainly exciting results await us in the decades to come.  

\section{Detecting Exoplanets}

The first theory that successfully described the motion of planets around the Sun was proposed by the German mathematician Johannes Kepler in his ``Laws of Planetary Motion''\footnote{He did not describe them as laws at the time of publication.}. These 3 laws explained the observations of several planets 	recorded by Danish astronomer Tycho Brahe with unprecedented accuracy. 

\begin{figure}[ht]
\begin{center}
\subfigure[The first law of planetary motion states that the orbits of planets are ellipses with the Sun at one of its foci.]{
   \includegraphics[width=0.4\linewidth]{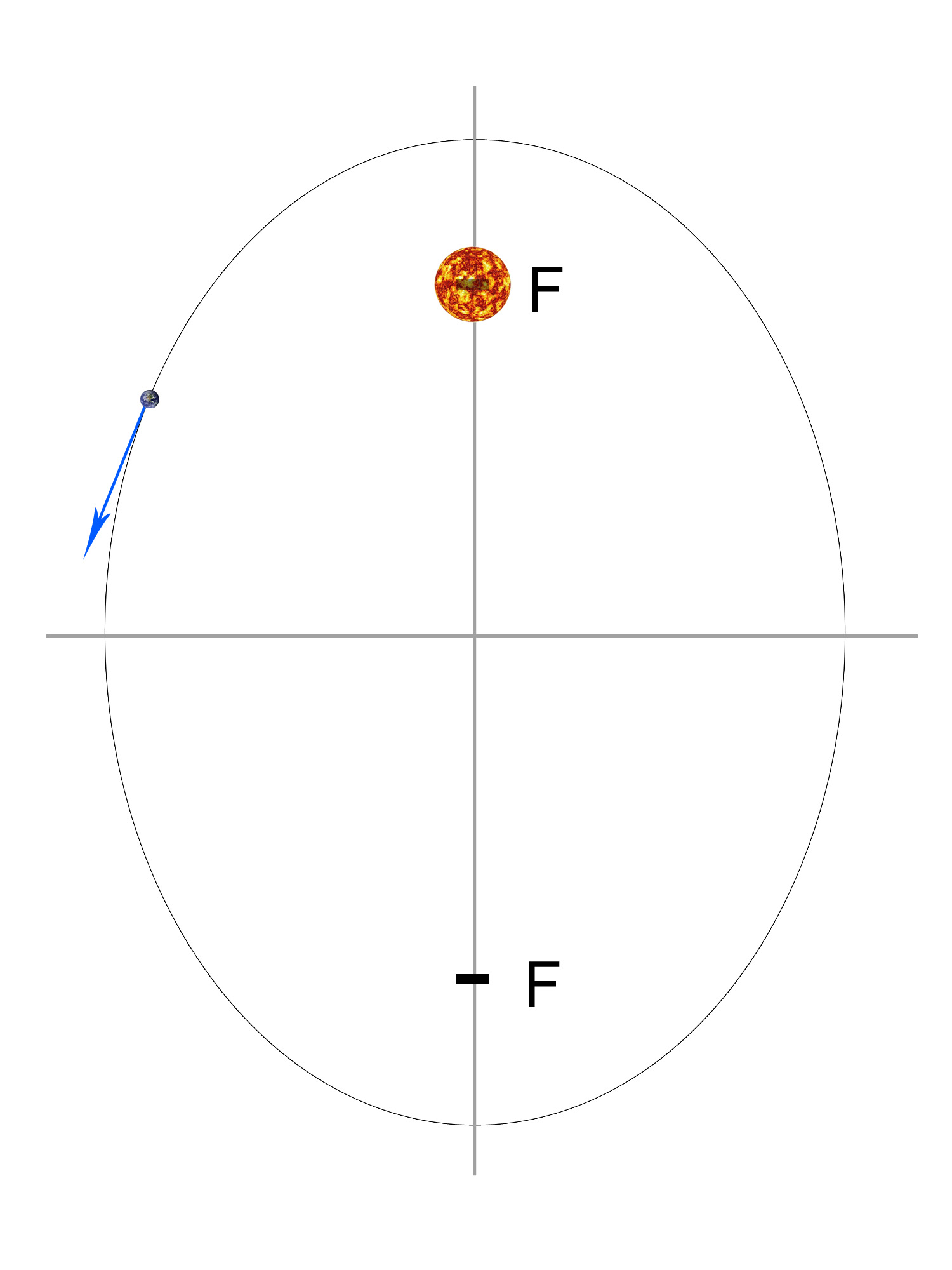}
   \label{fig:kepler_1}
 }\hfill
 \subfigure[The second law of planetary motion states that a line joining the planet and the Sun sweeps equal areas in equal periods.]{
   \includegraphics[width=0.4\linewidth]{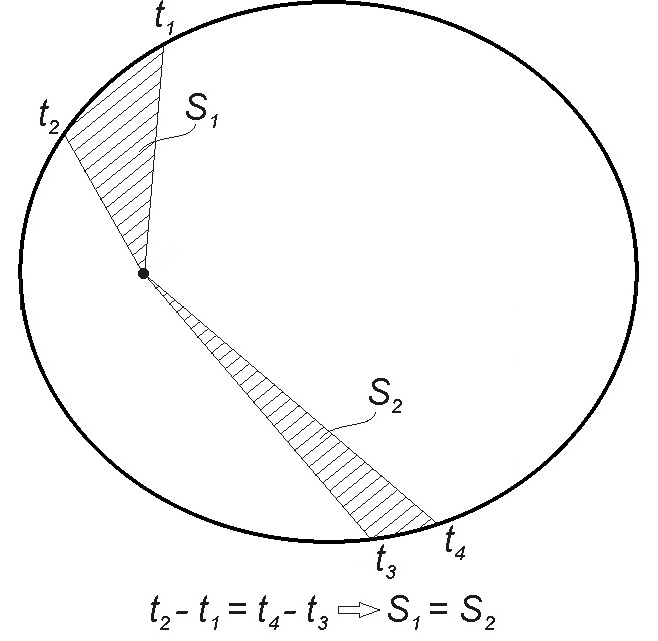}
   \label{fig:kepler_2}
 }
	\label{fig:kepler_12}
	\caption{Kepler's laws of planetary motion~\citep{voelkel_composition_2001}.}

\end{center}
\end{figure}

The first law removed the idea of circular orbits describing the motion of planets as elliptical and placing the Sun at one of its foci, see Figure~\ref{fig:kepler_1}. The second law states that the line connecting the Sun with the planet sweeps equal eras in equal periods of time, see Figure~\ref{fig:kepler_2}. The first 2 laws were published in ``Astronomia Nova'' in 1609~\citep{voelkel_composition_2001}. In 1619, the third law was published in ``Harmonices Mundi'' and it relates the cube of the semi-major axis of the orbit to the square of the orbital period~\citep{holton_physics_2001}. 

 \hspace*{\fill}
 \hspace*{\fill}
%
%

Planetary orbits around a star can also be understood in the context of a two body problem. In this case, the analysis is presented by the study of the mutual gravitational attraction. The solution was initially presented by~\cite{newton_philosophiae_1687}. It became clear that Kepler's laws of planetary motion were a natural consequence of Newton's universal law of gravitation,
\begin{equation}
	\label{eq:G}
	\mathbf{F}=G\frac{m_1m_2}{r^2} \mathbf{\hat{r}},
\end{equation} 
where $F$ is the force that arises from the gravitational interaction between bodies of masses $m_1$ and $m_2$ separated by the distance $r$. The universal gravitational constant $G=6.67260\times10^{-11}$N\,m$^2$\,kg$^{-2}$ , sets the scale of the force produced by this interaction. 

One of Newton's many contributions was to show that elliptical orbits arise from Equation~\ref{eq:G}. This derivation is included in Appendix~\ref{App:RV} and the result is expressed as the planet's velocity, $v$, as a function of its distance to the host star, $r$:

\begin{equation}
	\label{eq:V}
		\displaystyle	 v(r)=\pm\sqrt{G(m_1+m_2)\left(\frac{2}{r}-\frac{1}{a}\right)},
\end{equation} 

where $a$ is the semi-major axis of the orbital ellipse. An example of the application of Equation~\ref{eq:V} is the change of linear velocity of Earth as it changes distance from the Sun in its elliptical orbit.



\section{Detection Methods}

\begin{figure}[ht]
	\begin{center}
		\includegraphics[width=0.9\textwidth]{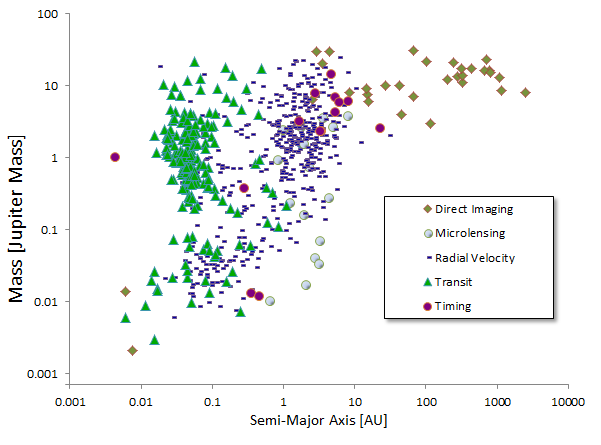}
	\end{center}
	\caption{Exoplanets discovered by method. The vertical axis shows the mass of the planet and the horizontal axis represents the distance between the planet and the star.}
	\label{fig:PerMass}
\end{figure}

The gravitational interaction between a planet and its host star creates periodical spatial displacements that can be spanned by three components. From a given point of view we can choose these three components to be the line of sight and its two perpendicular components, which are also perpendicular to each-other. The analysis of this information combined with the luminosity of the star form the essence of most exoplanetary search methods. Technologies for detecting exoplanets have only matured to produce quantitative results in the last 20 years. Several different methods have been in use, yielding different degrees of success, see Figure~\ref{fig:PerMass}.


\subsection{Radial Velocity}

This method makes use of the line of sight displacement of the host star to look for a companion. Because a single spatial dimension is being analysed, only the minimum mass of the companion can be calculated. The tilt of the plane of the orbit is unknown and this creates the potential discrepancies between the measured quantities and the object's actual values. Despite its limitations, this method has produced most of the results thus far.

The radial velocity of a body in a Keplerian bound orbit\footnote{A stable orbit that can be understood under Kepler's laws of planetary motion.} can be expressed in terms of the properties of the system. This is a useful result as it is the radial velocity of the host star what we can measure and its full derivation can be found in Appendix~\ref{App:RV}.

The radial velocity equation is:

\begin{equation}
	\displaystyle RV=\sqrt{\frac{G}{(m_1+m_2)a(1-e^2)}}m_2\sin \theta,
	\label{eq:RV}
\end{equation}
\vspace{5 mm}

where $RV$ is the radial velocity, $m_1$ and $m_2$ are the masses of the two bodies, $a$ is the semi-major axis, $e$ is the eccentricity of the orbit and $\theta$ the angle between the line of sight and the plane of the orbit. Equation~\ref{eq:RV} can be written in a more convenient way using dimensions of the solar system and simplifying $(m_1+m_2)\approx m_1$ for a star-planet configuration:
\vspace{5 mm}
\begin{equation}
	\boxed{\displaystyle	RV=28.4329  \left(\frac{m_1}{\msun}\right)^{-\nicefrac{1}{2}} \frac{m_2}{M_{Jup}} \left(\frac{a}{1 \text{AU}}\right)^{-\nicefrac{1}{2}} \frac{\sin \theta}{\sqrt{(1 -e^2)}} \text{ms}^{-1}}
\end{equation}
\vspace{5 mm}

or in terms of the orbital period $P$,
\vspace{5 mm}
\begin{equation}
	\boxed{\displaystyle	RV=28.4329 \left(\frac{m_1}{\msun}\right)^{-\nicefrac{2}{3}} \frac{m_2}{M_{Jup}} \left(\frac{P}{1 \text{yr}}\right)^{-\nicefrac{1}{3}} \frac{\sin \theta}{\sqrt{(1 -e^2)}} \text{ms}^{-1}}
	\label{eq:RV_P}
\end{equation}
\vspace{5 mm}

where $m_2$ is the mass of the planet, $m_1$ is the mass of the star, $M_{Jup}$ is the mass of Jupiter and $\msun$ the solar mass. 



\subsection{Transits}
The alignment of two celestial bodies receives different names based on the proportional sizes of the objects and distance between them. When they subtend a similar angular size, from a given point of view, it is called an eclipse. The Sun and the Moon are familiar examples of objects that subtend a similar angle in the sky. When the sizes are different, two options arise. The small object passing in front of the large one, covering only a small portion of the disk, is called a transit or primary eclipse. The opposite alignment is an occultation or secondary eclipse. The transit technique requires a particular alignment of the system with the line of sight, limiting the stars that are potentially good candidates for it. Large surveys are usually required to increase the statistical chances of finding the right system. For a random orientation, and assuming a planet much smaller than its host star, the probability, $p$, to transit is
\begin{equation}
	p=\frac{R}{a},
	\label{eq:TRProb}
\end{equation}
where $R$ is the radius of the star and a the semi-major axis. For a Sun-like star, the probability of a planet transiting at 1 AU is (1/125). Transits at shorter distances are much more common, and also have shorter periods, making them better candidates for detection. The Kepler mission~\citep{middour_kepler_2010} currently scans a fixed region of 115 square degrees that includes 100,000 stars every 30 minutes. It is a mission designed to maximize the opportunities to observe exoplanets by the transit method. In over 2 years, it has discovered 61 exoplanets and found over 2300 candidates. 

\subsection{Other Methods} 

\subsubsection{Astrometry}
Unlike the radial velocity method, astrometry is based on the components of motion in the plane of the sky. There are certain advantages that astrometry provides that can improve the data obtained by other methods. The true mass of the planet can be obtained as two components of the orbital motion are known, hence the inclination of the system can be derived. This also applies to multiple planet systems as the high eccentricity observed can many times describe non-coplanar orbits. The independence of spectral type allows for any type of star to be observed producing an unbiased characterization of planetary properties as a function of stellar type. Astrometric signals increase linearly with the semi-major axis, so, in principle, it is a method that produces its best results from multiple planet systems whose more massive members are contained in the inner orbits.

\subsubsection{Timing}
Pulsars~\citep{wolszczan_planetary_1992} and stellar oscillations~\citep{silvotti_giant_2007} provide very precise temporal information. Periodic pulse deviations from the expected frequency can provide detailed information on the radial motion of the star. This can be the signature of an unseen companion. A triple system orbits pulsar PSR B1257+12~\citep{wolszczan_25.3_2000}, where the innermost planet is only twice the mass of the Moon. Pulsar timing is so precise, that in principle it could allow us to detect masses as small as a large asteroid. The detection is based on the radial motion of the star, so only mass and orbital elements can be derived.
  
\subsubsection{Gravitational Microlensing}
In the presence of mass, spacetime curves bending light as a consequence~\citep{einstein_erklarung_1915}. When the alignment of two stars coincides with the line of sight, the light from the background star is curved. The foreground star acts as a lens. The characteristics of this optical arrangement depend on the mutual distance, the luminosity of the background star and the mass of the lens star. By analysing the light curve, information about the foreground star can be obtained. Discrepancies in the expected mass of the foreground star can be evidence of a companion. In principle this technique can be sensitive enough to detect Earth mass planets, sensitivity only exceeded by the timing technique. The clear disadvantage is that this is a one-time event and any further observations can only be obtained using other techniques. 
  
\subsubsection{Direct Imaging}
The possibility to take a snapshot of an exoplanet requires the ability to spatially separate the planet and the star. The difficulty in this technique resides in the large difference in brightness between the star and the planet. This is currently limited to big, bright, young or massive sub stellar objects far away from the host star~\citep{marois_direct_2008, kalas_optical_2008}. To be able to detect solar system like planets, the sensitivity to discern the brightness between the two objects needs to improve about 100 times from the current technological limits. Despite these difficulties, direct imaging is responsible for 12 confirmed direct planet detections. It also opens the possibility to detect planets in formation~\citep{kraus_lkca_2012}, or even protoplanetary disks, the disks surrounding stars during their early stages. This broadens the scope of science that can be obtained from exoplanetary research, increasing the understanding of the formation of planetary systems from the earliest stages.

When stars deplete their hydrogen fuel, they enter the first red giant phase~\citep{iben_low-mass_1968}. Planetary systems are disrupted during this process. Inner planets find themselves engulfed in the growing outer shells of their host star, and outer planets become disrupted under these changing conditions. This phase of changes is followed by a quieter helium burning stage known as the red clump phase~\citep{sweigart_evolutionary_1978}. Stars in this phase tend to have between 1.0 and 2.5 solar masses. There is a gap in our current understanding on survival rate of planetary systems over this stage and how they affect their host star. It is the goal of my project to develop the technologies that will quantitatively fill in this gap. 


	\chapter{Spectroscopy}

\section{Concepts}
\subsection{Astronomical Spectroscopy}

One of the most successful ways to extract information from light is to spread its energy into its constituting wavelengths to analyse them independently. There are several ways to achieve this. A filter is an intuitive, and certainly effective, way to isolate a range of wavelengths from the full spectrum. Inserting the filter in the beam of incoming light is a way to pick only a fraction of the full spectrum, potentially revealing information on the source. Nonetheless, trying to capture several wavelengths independently in the same session can be ineffective by this method and a different approach is needed. A continuous spread of a range of wavelengths presenting the relative intensities can be captured by the use of dispersive optical components. These type of elements use different properties of light to alter its path, effectively allowing us to discern between the different energy levels of neighbouring wavelengths. There are different ways to achieve this, the most relevant to this project are presented below.

\subsection{Prisms}

 The \textbf{refractive index} of a material is represented by the ratio of the speed of light in vacuum~($c$) to the speed of light through the material~($v$). It is a function of the wavelength of the beam travelling through the medium. 
	
\begin{equation}
	n(\lambda)=\frac{c}{v(\lambda)}
\label{eq:Refractive_Index}
\end{equation}  

	The change in direction of a beam of light when crossing through refractive media with different indices of refraction is represented by \textbf{Snell's law}. 
	
\begin{equation}
	n_1 \sin{\theta_i} = n_2 \sin{\theta_r}
\label{eq:Snell}
\end{equation}   	 

where $\theta_i$ and $\theta_r$ are the subtended angles of the beam as it crosses media with refractive indices $n_1$ and $n_2$ respectively. This effect is used in prisms to disperse the different colours that form light. 

\begin{figure}
	\begin{center}
		\includegraphics[width=0.5\linewidth]{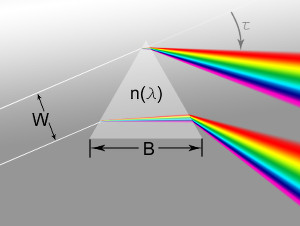}
	\end{center}
	\caption{The angle of refraction is dependent on wavelength making a prism a dispersive optical component.}
	\label{fig:prism}
\end{figure}

\begin{figure}[ht]
\centering

\subfigure{
   \includegraphics[width=0.45\linewidth]{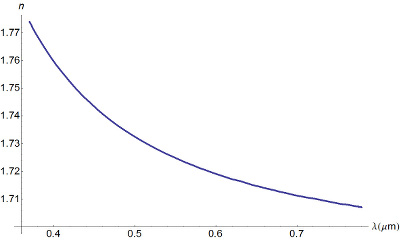}
   \label{fig:N-KZFS8}
 }
 \subfigure{
   \includegraphics[width=0.45\linewidth]{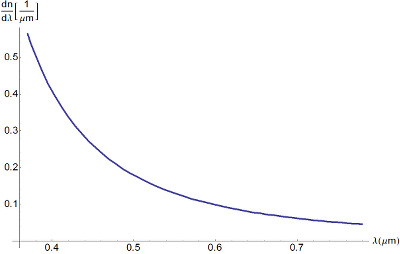}
   \label{fig:Disp_N-KZFS8}
 }
	\label{fig:Disp_N-KZFS8_2}
	\caption{Refractive index and dispersion of N-KZFS8 as a function of wavelength.}
\end{figure}

An empirical approach to the determination of the refractive index of a transparent material as a function of wavelength is the \textbf{Sellmeier Equation}. In its general, temperature independent form,

\begin{equation}
	n^2(\lambda) = 1 + \frac{B_1 \lambda^2 }{ \lambda^2 - C_1} + \frac{B_2 \lambda^2 }{ \lambda^2 - C_2} + \frac{B_3 \lambda^2 }{ \lambda^2 - C_3},
\end{equation}

provides a refractive index as a function of wavelength.  The $B$ and $C$ parameters depend on the material used and it is provided by the manufacturer. The glass of the prism used in the RHEA spectrograph is N-KZFS8, see Figure~\ref{fig:Disp_N-KZFS8}

The \textbf{angular dispersion} is a measurement of the change of the resulting angle as we move through the wavelengths.

\begin{equation}
	\frac{d\tau}{d\lambda}=\frac{B}{W}\frac{dn}{d\lambda}
\end{equation}

where $\tau$ is the refraction angle of the beam, $B$ is the base of the prism and $W$ the width of the beam, see Figure~\ref{fig:prism}.

The \textbf{spectral resolution}, $R$, produced by a prism is related to rate of change of refractive index and the length of the prism base \citep{schroeder_astronomical_2000}. In the diffraction-limited case that is:

\begin{equation}
	R=B\frac{dn}{d\lambda}.
\end{equation}

As an individual optical dispersing element in a spectrograph, a prism would require impractical dimensions to reach the resolving power of $\sim$50,000 across the full optical range.

\subsection{Gratings}

An alternative dispersing element is the grating. The diffraction of wave fronts reflecting at different angles from a grating surface will form an interference pattern peaking in intensity where the waves produce constructive interference. This process is what allows a grating to spread a monochromatic beam into different peaks or orders. 

\begin{figure}[H]
	\begin{center}
		\setlength\fboxsep{0pt}
		\setlength\fboxrule{0pt}
		\fbox{\includegraphics[width=0.5\linewidth]{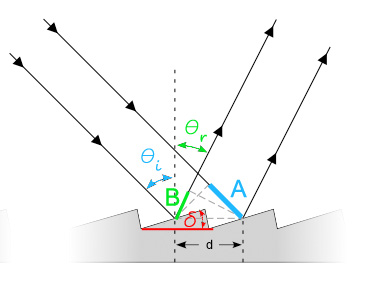}}
	\end{center}
	\caption{A grating produces constructive interference when the path difference (A-B) is equal to an integer number of wavelenghts.}
	\label{fig:echelle-grating}
\end{figure}

The geometry of two parallel beams of hitting a grating is represented in Figure~\ref{fig:echelle-grating}. It can be noted that

\begin{equation}
	A=d \sin {\theta}_i 
\label{eq:A}
\end{equation} 

\begin{equation}
	B=d \sin {\theta}_r.
\label{eq:B}
\end{equation} 

The path difference has to equal an integer number of wavelength to create constructive interference, hence we are looking for the angles where

\begin{equation}
	\begin{array}{rcl}
		A-B & = & \displaystyle	 (d \sin {\theta}_i - d \sin {\theta}_r) \\
			& = & \displaystyle	 d(\sin {\theta}_i - \sin {\theta}_r) \\
			& = &  \displaystyle	n\lambda	
	\end{array}
\end{equation}

producing the grating equation

\begin{equation}
	n\lambda= d(\sin {\theta}_i - \sin {\theta}_r).
	\label{eq:Grating_Eq.}
\end{equation}

\subsubsection{Intensity distribution}

The grating equation does not describe how the energy is distributed. The efficiency of a grating for a given wavelength is a combination of the interference pattern that arises from successive grooves, and the blaze function, that shapes the diffraction of a single groove. The action of tilting the grating so that each blaze gives specular reflection is called blazing. Effectively it matches the peaks of the interference pattern with the peak of the blaze function maximising the efficiency of the grating. 

The wavelength that peaks at a given order is called the blaze wavelength and is found by 

\begin{equation}
	\lambda_b=\frac{2d\sin\delta\cos(\theta_i-\delta)}{n}
	\label{eq:BlazeAng}
\end{equation}

where $\delta$ is the blaze angle.  A special case for Equation~\ref{eq:BlazeAng} is when the angle of the incident beam is the same than the blaze angle ($\theta_i=\delta$), this is called Littrow configuration and is the approach adopted for this project, in this case Equation~\ref{eq:BlazeAng} becomes 

\begin{equation}
	\lambda_{b(Litt)}=2d\sin\delta
	\label{eq:Littrow}
\end{equation}

\subsubsection{Echelle Gratings}

Increasing the angle of incidence or the blaze angle leads to higher resolution. However there is a limit on the steepness that the ruling is allowed to have before overlapping occurs. This type of grating with steep and coarse ruling are called echelle gratings. They are classified by the R-number which indicates the tangent of the blaze angle. R2 is the most common type of echelle grating, $\delta\approx$63.4349$\,^{\circ}$, and it is the type used in this project.

This type of grating has to operate at high orders producing a collection of short spectral orders due to the limited free spectral range. To separate the overlapping orders an extra dispersing optical component is needed. In the RHEA spectrograph, a prism is placed in the output beam from the grating acting as a cross-disperser effectively separating the otherwise overlapping orders. It operates in a perpendicular direction than the grating producing a 2-dimensional dispersion making a more efficient use of the CCD detector. 

\subsubsection{Spectral Resolution}

A critical value that characterises a grating in an optical arrangement is the angular dispersion it produces. This value expresses the rate of change of wavelength as we sweep the output angles from the grating. Rearranging Equation~\ref{eq:Grating_Eq.} we find  

\begin{equation}
	\theta_r = \sin^{-1}(\frac{n\lambda}{d}+\sin{\theta}_i).
\end{equation}

and taking the derivative 

\begin{equation}
	\begin{array}{rcl}
		\displaystyle	\frac{d\theta_r}{d\lambda} & = &  \displaystyle	 \frac{n}{d}(1-(\frac{n\lambda}{d}+\sin{\theta}_i)^2)^{-\nicefrac{1}{2}} \\[2.5ex]
    						& = &  \displaystyle	\frac{n}{d}(1-sin^2{\theta}_r)^{-\nicefrac{1}{2}} \\[2.5ex]
    						& = &  \displaystyle	\frac{n}{d \cos{\theta}_r} \\[2.5ex]
    						& = &  \displaystyle	\frac{\sin{\theta}_i-\sin{\theta}_r}{\lambda \cos{\theta}_r} \\[2.5ex]
    						& = & \displaystyle	 \frac{2}{\lambda} \tan{\theta}_r
	\end{array}
\end{equation}

rearranging

\begin{equation}
	\frac{\lambda}{\Delta\lambda} = \frac{2}{\Delta\theta_r} \tan{\theta}_r = \text{R}
	\label{eq:dtheta}
\end{equation}

and in the diffraction-limited case 

\begin{equation}
	 \text{R} = \frac{\lambda}{D} 
\end{equation}

\subsection{Telescopes}

\subsubsection*{Focal ratio}
The ratio of the clear aperture of the telescope to the focal length, the distance between the first corrective surface and the focus, is the focal ratio. It is a measurement of the steepness that the beam will need in order to find focus at a given distance. 
\begin{equation}
	\rm focal ratio(\nicefrac{f}{\#})=\frac{F}{D}
\end{equation}

\subsubsection*{Plate Scale}
The ratio of angle per distance that a telescope will produce on its focal plane is called the plate scale.

\begin{equation}
	\rm PlateScale=\frac{1}{D\times\nicefrac{f}{\#}}\text{rad mm}^{-1}=\frac{206265}{D\times\nicefrac{f}{\#}}^{\prime\prime}\text{mm}^{-1}
\end{equation}

\subsubsection*{Resolving Power}
\begin{figure}
	\begin{center}
		\includegraphics[width=0.5\linewidth]{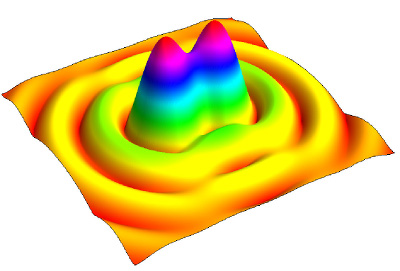}
	\end{center}
	\caption{The distance between two adjacent Airy discs that can be individually identified is the resolving power of the telescope.}
	\label{fig:Airy}
\end{figure}

A concept closely related to the plate scale is the resolving power. When a point source is viewed at the focal plane, the best we can resolve is a diffraction pattern (e.g. an Airy disc) of finite size, see Figure~\ref{fig:Airy}. A telescope under this operation is said to be diffraction limited. The resolving power of a telescope is the capacity to resolve two adjacent Airy discs. It is expressed in terms of the angular separation the sources need to have in order to produce such an image a the focal plane. A measurement of this quantity is the Dawes' limit. 
\begin{equation}
	R=\frac{116}{D}^{\prime\prime}
\end{equation}

where $D$ is the telescope's main aperture in millimetres. The intuitive interpretation of this value is that two source objects can be as close as $R$ arcseconds from each other and they will still be identified as two separate objects, assuming diffraction limited resolution is possible.

\section{Radial Velocity Calibration}

Precision radial velocity can yield information about properties of a star and its orbiting bodies that were unknown before this technique was used. Over the last 20 years a wide range of techniques have improved the spectral and temporal resolution that these measurements can yield, exposing internal stellar processes and unveiling exoplanets.

\cite{vogel_spectrographic_1892} demonstrated that stars in motion along the line of sight would exhibit a change in color. Earlier work had tried to apply Doppler's theory to the motion of the stars but technology was not ready yet. It was the improvements in the spectrograph at the Royal Observatory in Postdam, Germany, that proved this effect conclusively. Since then, Doppler shift has been used largely in star velocity measurements. \cite{wilson_general_1953} published the \emph{General Catalogue of Stellar Radial Velocities} containing data on 15,000 stars. However, typical radial velocity precisions were of the order of $\approx$1kms$^{-1}$, not yet enough for planet detection. 

The use of a stable secondary source to improve resolution was proposed initially in 1973~\citep{griffin_possibility_1973}. The absorption lines produced by the nearly stationary Earth's atmosphere could be used as a reference to increase accuracy. Although using other secondary sources to produce a reference spectrum provides benefits, Griffin outlined several advantages of using telluric\footnote{The spectral lines produced by the Earth's atmosphere.} lines instead; the optical path is the same in the star and the reference, the reference has an absorption spectrum, which makes it comparable to the star's and it is always `turned on'. Despite not reaching the 10$\ms$ expected, these concepts led to great improvements in radial velocity precision in the following years. 

\cite{campbell_precision_1979} introduced a method for inserting a hydrogen fluoride cell in the light path of a coud\'{e} spectra\footnote{Telescopes working at coud\'{e} focus are designed to keep the focal plane at a fixed location, despite its orientation. This allows for large instruments to be used without weight restrictions.}. This approach increased the achievable accuracy to 15$\ms$\,and led to a 12 year project that monitored 17 main-sequence stars\footnote{Middle-aged stars still burning hydrogen in their cores.} using the \emph{Canada France Hawaii Telescope}~(CFHT)~\citep{walker_search_1995}. The radial velocity precision achieved a new milestone with the use of Iodine cells as a reference, leading to an accuracy of 3$\ms$~\citep{butler_attaining_1996}.

\begin{figure}
\centering

\subfigure[CCD frame of a CORALIE stellar exposure with its simultaneous thorium reference.]{
   \includegraphics[width=0.4\textwidth]{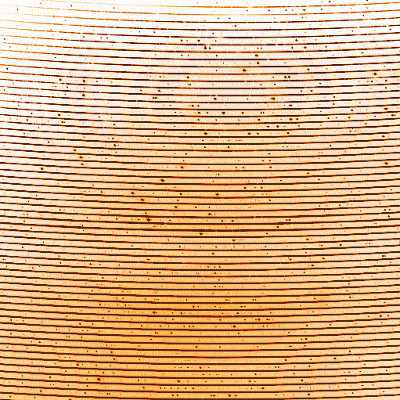}
   \label{fig:CORALIESpec}
 }
 \subfigure[4000$\times$4000 pixel frame of a K0V star showing 70 orders spanning the visible spectrum from HARPS.]{
   \includegraphics[width=0.4\textwidth]{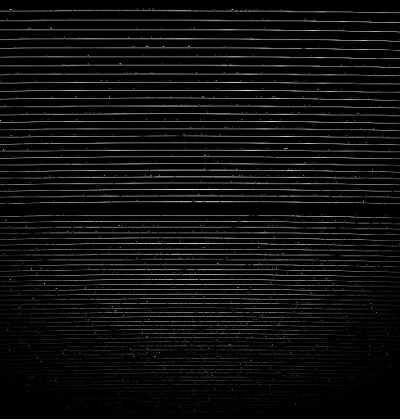}
   \label{fig:HARPSSpec}
 }
	\caption{Emission lines from thorium argon are visible in both images between the orders of the 
stellar spectrum.}
	\label{fig:COREALIE_HARPS}
\end{figure}

In 1998, the European Southern Observatory (ESO) issued a proposal to develop an instrument that could reach a 1$\ms$\,precision dedicated to the search for exoplanets. A consortium formed by several organizations across Europe developed the \emph{High Accuracy Radial velocity Planet Searcher}~(HARPS)~\citep{mayor_setting_2003}. After successful tests in early 2003, HARPS commenced operations at La Silla 3.6 m telescope in Chile as the successor of the CORALIE spectrograph~\citep{queloz_coralie_2001}. HARPS is a fibre fed, cross-dispersed echelle spectrograph. It is fed by two different optical fibres, one carries the light from the star targeted by the telescope, and the other carries a reference spectrum produced by a Torium-Argon~(ThAr) lamp. An example of a similar double spectra from CORALIE is shown in Figure~\ref{fig:COREALIE_HARPS}.

\section{Stellar Oscillations}

With the increase in radial velocity precision, new phenomena arises in the finer details that become visible. Stellar oscillations can cause an apparent shift on radial velocities~\citep{jimenez_radial_1986,deming_apparent_1987}. The understanding of these oscillations is fundamental to the correct calibration of radial velocity measurements as their signature can be of similar amplitude than the velocity induced by orbiting planets. 

\cite{mcmillan_radial_1993} observed the Doppler shift of the solar spectrum over a period of 5 years. Using the sunlit surface of the moon, the spectrum of the Sun integrated over its surface could be measured. This provides results similar to the ones that could be obtained by observing the Sun at stellar distances. The Doppler shifts found varied less than $\pm$4$\ms$. Astroseismological activity, similar to that observed in the Sun, has been detected in several Sun-like stars~\citep{martic_evidence_1999,bedding_evidence_2001,bouchy_p-mode_2001}. The key to discriminate the stellar oscillations from the signature of exoplanets, is to gain understanding on how solar oscillations scale with stellar properties. 

\begin{figure}
	\begin{center}
		\includegraphics[width=300px]{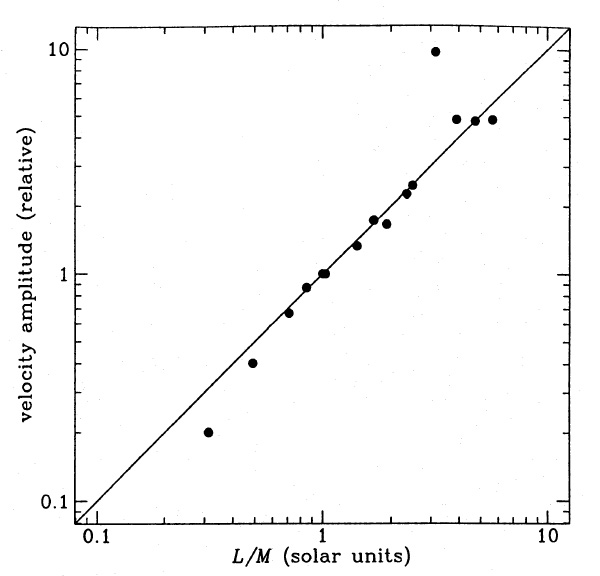}
	\end{center}
	\caption{Velocity amplitude versus light-to-mass ratio for solar-like oscillations~\citep{kjeldsen_amplitudes_1995}.}
	\label{fig:L_to_M}
\end{figure}

From a large sample of oscillating stars, \cite{kjeldsen_amplitudes_1995} calculated the relation 

\begin{equation}
	{\left(\frac{\delta L}{L}\right)}_{bol}\propto {\frac{V_{osc}}{\sqrt{T_{eff}}}},
	\label{eq:AstrSeis}
\end{equation}

where $\frac{\delta L}{L}$ is the star's luminosity oscillation, $T_{eff}$ is the effective temperature and $V_{osc}$ is the observed velocity amplitude. This proportionality, combined with the oscillations measured from the Sun, yield a general relation

\begin{equation}
	V_{osc}\propto \frac{L}{M},
\end{equation}

that represents the relation between the observed velocity amplitude and the luminosity to mass ratio, $\frac{L}{M}$. Figure~\ref{fig:L_to_M} shows the relation measured from several stars expressed in units relative to the Sun. The velocity oscillations measured in the Sun are $\sim$0.255$\ms$\citep{libbrecht_advances_1991}.

In the case of giant stars of 1.0 to 2.5 $\msun$, the amplitude of these oscillations is of the same order as the signal shift produced by the presence of a companion.  To be able to discriminate between both, observations over several hours to a full night need to be undertaken. Large telescopes are impractical for this task due to their high demand and operational costs, this is one of the reasons for the limited data available in this range of stars. Small telescopes can achieve these results at a fraction of the cost of professional telescopes. 

Understanding stellar oscillations, permits the disentanglement of the planetary signals in the measured radial velocities. \cite{christensen-dalsgaard_physics_2004} related the period of sun-like stars to the square root of the mean density, 

\begin{equation}
	P=2 \pi \sqrt{\frac{R^3}{G M}}
\end{equation}

where $M$ and $R$ are the mass and radius of the star and $G$ is the gravitational constant. For red giants stars, the dominating oscillations have a period of 3 to 10 hours. Observations over several nights can allow us to average over several oscillations, reducing the error introduced by stellar oscillations bellow the maximum oscillations measured. As the number of exoplanets discovered increases, the trends in their characteristics challenge the model set by our Solar System~\citep{erskine_externally_2005}. High resolution dispersion is required to obtain the necessary precision for planet detection ($R=\frac{\lambda}{\delta\lambda}\sim10^{5}$)~\citep{vogt_hires:_1994,vogt_lick_1987,mayor_setting_2003}.

\section{The Use of Small Telescopes}

Stellar radial velocity observations include the noise produced by the intrinsic oscillations~\citep{brown_afoe:_1994, bedding_solar-like_2003}. The uncertainty introduced by the photon noise is

\begin{equation}
	\sigma_{RV}=\frac{c}{Q\sqrt{N_{e-}}},
\end{equation}

where $Q$ is the quality factor, $c$ the speed of light and N$_{e-}$ the total number of photoelectrons counted over the whole spectral range \citep{bouchy_fundamental_2001}. The quality factor, $Q$, is a function of the spectral type of the observed star and independent of the flux.

\begin{figure}
	\begin{center}
		\includegraphics[width=300px]{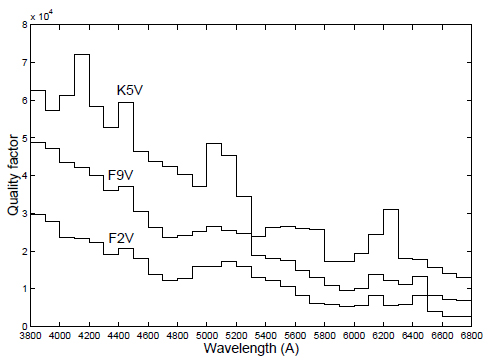}
	\end{center}
	\caption{Quality factor versus spectral range for a K5V, F9V and F2V star.~\citep{bouchy_fundamental_2001}.}
	\label{fig:QvsLambda}
\end{figure}

Radial velocity measurements rely on the presence of spectral features in the emitted spectrum. As most of the emitted star radiation ranges from the ultra-violet to the mid infra-red, ground based radial velocity measurements can only focus on the small window formed by the visible and near infra-red spectrum. The available stellar spectral lines within this window will depend on several factors being temperature the most relevant. 
On hot stars, where $T>10000K$, there are no electron transitions in the existing atoms effectively rendering the spectrum window a continuous with no lines to use as reference for radial velocity measurements. In addition, hot stars tend to rotate, smearing even further any spectral line. In cool stars, $T<3500K$, spectral lines are densely packed due to complex molecules allowed at lower temperatures. These stars are intrinsically faint and peak their emission in the infra-red, adding technical difficulties due to low SNR.
This constrains the range of stars that are ideal for radial velocity measurements. The stars that stand out as ideal are in the range between 0.1 to 1.5 $\msun$ during the main stage of their lives, called the main sequence. Most efforts so far have been focused on these type of stars. An exception to this limited regime, is the Red Giants and their metal rich counterparts, the Red Clumps. In astronomy, metals is an umbrella name for all elements heavier than Helium. These cool and slowly rotating stars are part of the candidates that show strong emission lines observable by ground based spectrographs.

A key aspect of radial velocity measurements is to understand how precision depends on the shape of the spectral lines detected. There are three characteristics that will determine the precision achieved: the ratio of useful to background information on the spectrum\footnote{the signal-to-noise ratio or SNR}, the depth of the spectral line being analysed, and its width. All these features can be expressed by 

\begin{equation}
	\sigma_{RV}\approx \frac{\sqrt{FWHM}}{C \cdot SNR},
	\label{eq:sigma1}
\end{equation} 

where FWHM is the full width half maximum of the line, $C$ is its contrast or depth and SNR is the signal-to-noise ratio. It becomes clear that a large FWHM, limited either by the source or by the instrument, will compromise the achievable precision. The increase in the rotational velocity of the star, or decrease in the resolution of the instrument, simultaneously increases the FWHM and decreases the contrast, as the total width should be conserved. This means that $C\approx\nicefrac{1}{FWHM}$ so effectively the RV precision degrades as $FWHM^{\nicefrac{3}{2}}$. 
	\chapter{Instrumentation}

The motive behind the RHEA spectrograph is the construction of the simplest spectrograph that would reach the level of precision necessary for the detection of exoplanets. We use a ``from-the-ground-up'' approach where an initial spectrograph setup is attempted and failing points are noted to be improved in subsequent versions. This approach ensures that only the minimal configuration becomes the standard in the final version in an attempt to reduce production costs and increasing potential interest for replicability.
This chapter presents the design of the current version of the RHEA spectrograph. 

\section{Optical Arrangement}

The full instrument setup includes the RHEA spectrograph and the support systems that feed the light to make the relevant measurements. There are several requirements that need to be successfully addressed in order to provide illumination in an efficient and reliable way. These instruments, containing both hardware and software components, are currently under development with different degrees of completion. 
The dome and slit control system keeps the dome is pointing in the right direction. The slits can be automatically opened or closed if needed. The weather information system provides feedback so that no instrument is exposed to rain and measures cloud density. The pointing and tracking systems ensure that the telescope is pointing in the right direction. The wide tracking loop uses a wide field camera mounted on the telescope and provides feedback on the orientation and tracking. The injection tracking loop is part of the fibre feed and monitors the correct alignment of the targeted star at the fibre entrance. The temperature stabilization system keeps the spectrograph at a constant temperature to ensure the that no thermal variations affect the calibration.

\section{The RHEA Spectrograph}

\subsection{Key Features}

The RHEA spectrograph is a high resolving power instrument, R$\sim$50000. It operates over a wavelength range between 400 nm and 795 nm. It is adapted to work with a 0.4m telescope working at F/10 focal ratio. The collimator lens operates at F/8, has a focal length of 200mm and a diameter of 25mm. The pupil of the system is defined by the prism and it's 9mm in aperture. The main dispersing component is an R2 echelle grating with a $\sim$63.43$\,^{\circ}$ blaze angle and 31.6 Grooves/mm. The prism acts as a cross disperser, it's made of N-KZFS8, has a 8mm base and a 30$\,^{\circ}$ apex angle. The sensor is a CCD Kodak KAF-8300 3326 $\times$ 2504 with 5.4 $\mu$m pixels. It includes a thermal stabilization system. The spectrograph is enclosed by a 5mm lightweight polystyrene foam and surrounded be a thermal insulator. It is fully constructed with off-the-shelf components. The camera shutter is the only moving part.

\subsection{Components}

 \begin{figure}[H]
\centering
\hspace*{\fill}
 \subfigure[The AC254-200-A collimating and camera \textbf{lens} has a focal length of 200mm and a diameter of 25mm. It's an achromat and operates at wavelength range of 400-700 nm] {
   \includegraphics[width=0.45\linewidth]{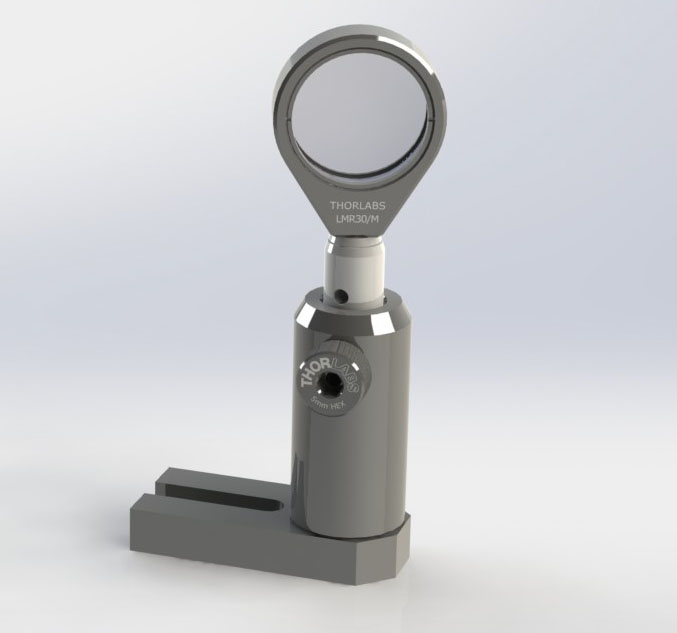}
 }\hfill
 \subfigure[The SBIG  STT-8300M \textbf{camera} has a Kodak KAF-8300 sensor with an array of 3326 $\times$ 2504 with 5.4 $\mu$m pixels.]{
   \includegraphics[width=0.45\linewidth]{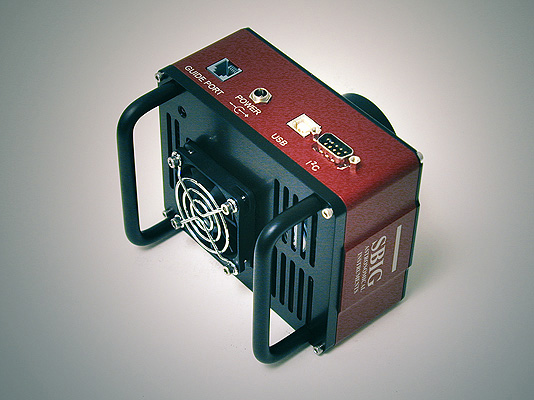}
 }
 \hspace*{\fill}
 \end{figure}

\begin{figure}[H]
\centering
\hspace*{\fill}
 \subfigure[The LM05XY/M \textbf{fibre optic attachment} is connected to a translating lens mount for $\phi\nicefrac{1}{2}''$ optics. It works with a sensitivity of 250$\mu m$/rev and it has a FC/PC connector ]{
   \includegraphics[width=0.4\linewidth]{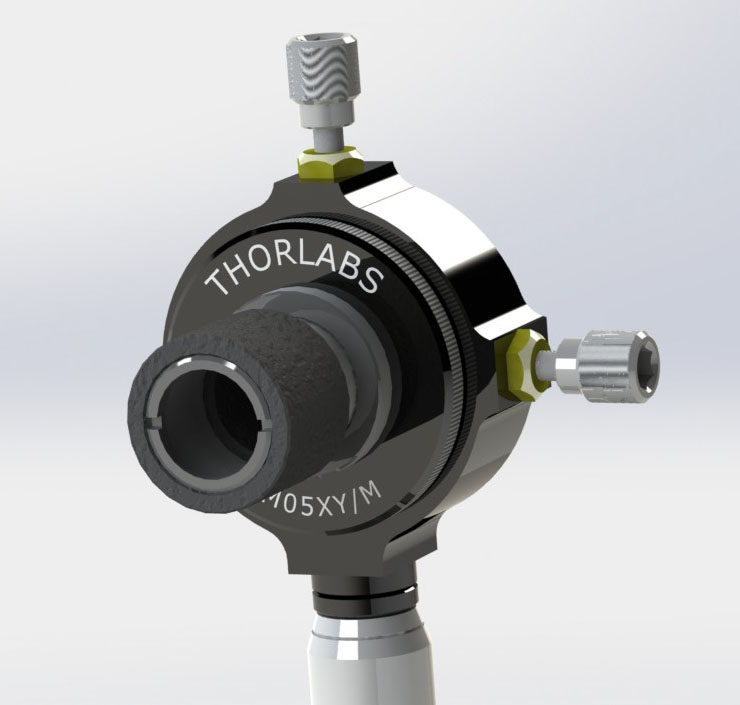}
 }\hfill
 \subfigure[ The PS873-A \textbf{prism} operates at a wavelength range of 350-700 nm has an apex angle of 30$\deg$. It's made of N-KZFS8 with a refractive index of 1.7249 @ 550nm]{
   \includegraphics[width=0.40\linewidth]{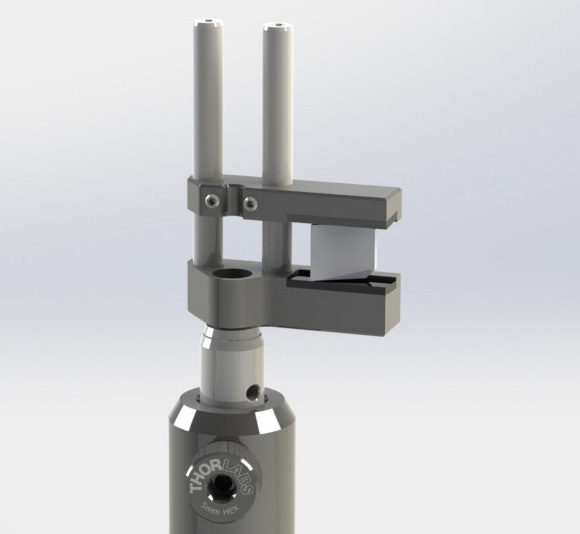}
 }
  \hspace*{\fill}
   \end{figure}

 \begin{figure}[H]
\centering 
   \hspace*{\fill}
 \subfigure[The GE2550-0363 \textbf{echelle grating} has 31.6 Grooves/mm and 63$\,^{\circ}$ blaze. The size is 25 $\times$ 50 $\times$ 9.5 mm]{
   \includegraphics[width=0.45\linewidth]{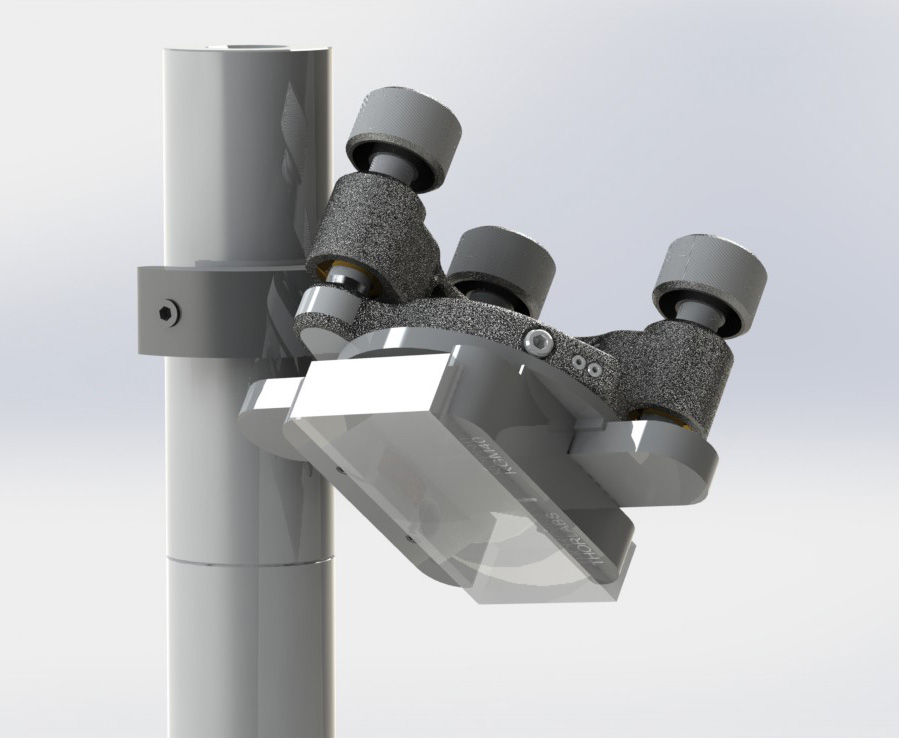}
 }\hfill
 \subfigure[The PFSQ10-03-F01 UV enhanced aluminium \textbf{mirror}is 25.4 $\times$ 25.4 mm in size]  {
   \includegraphics[width=0.45\linewidth]{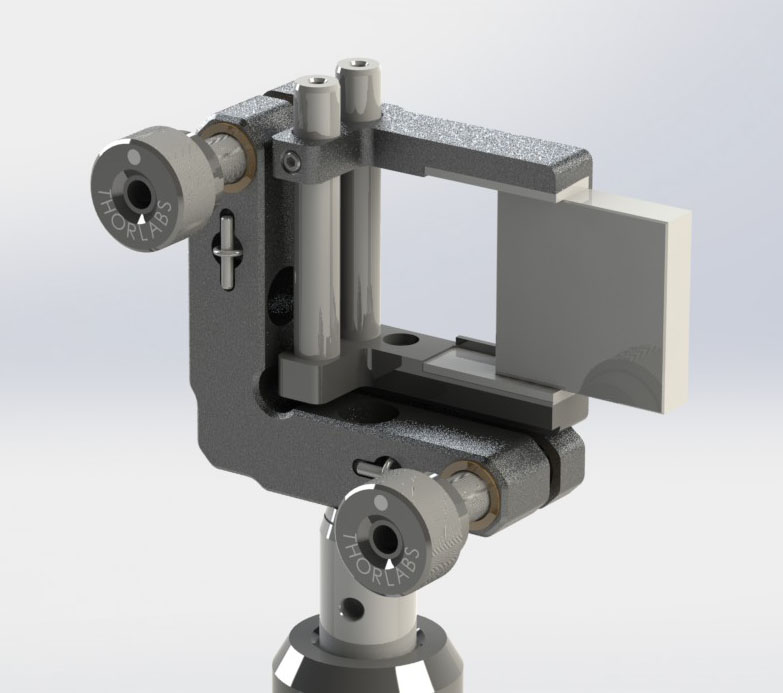}
 }
 \hspace*{\fill}
 \end{figure}

\subsection{Layout}

\begin{figure}[H]
	\begin{center}
		\includegraphics[width=0.9\linewidth]{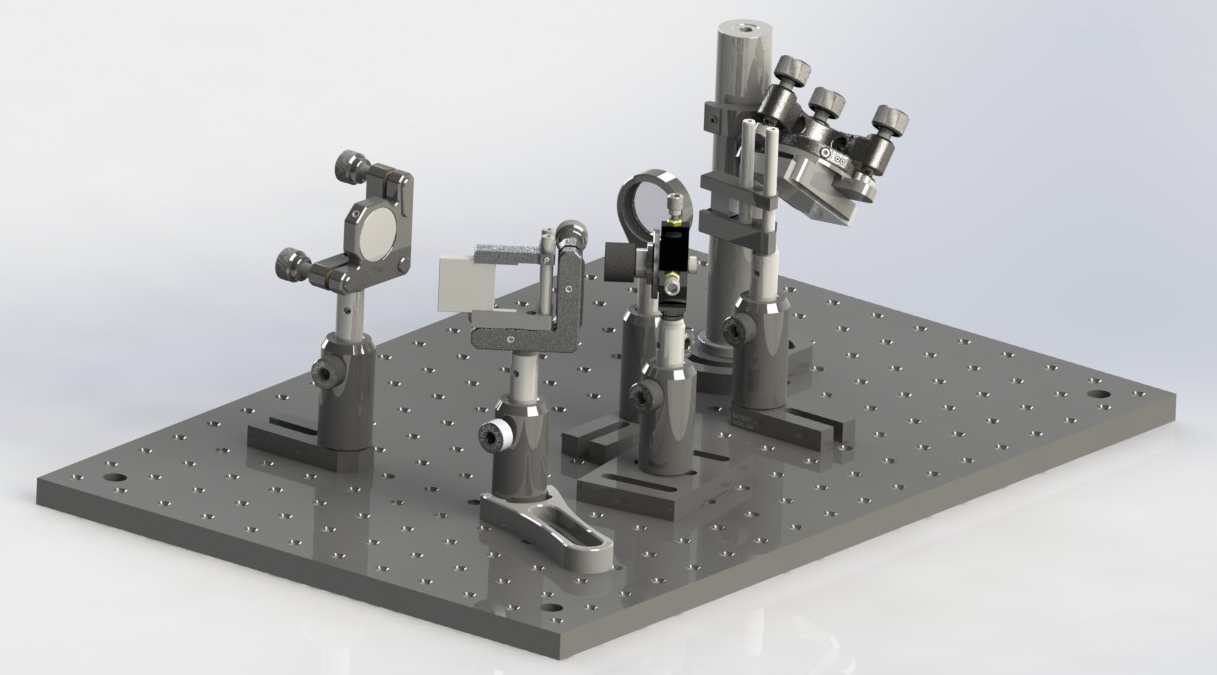}
	\end{center}
	\caption{The layout of the RHEA spectrograph.}
	\label{fig:SpecLayout}
\end{figure}

The spectrograph is designed keeping size and cost in mind. It relays the telescope by the use of a single mode fibre with a $\phi=\nicefrac{1}{2}'$ lens at the fibre attachment. The 3.5$\mu$m fibre core is relayed to create an 18.33$\mu$m size image that becomes the entrance slit. The beam is collimated by the 200mm focal length lens at f/8 and sent to the prism. The horizontally dispersed light is dispersed vertically by the grating placed at Littrow configuration. The returning light is dispersed once again by the prism for cross dispersion effectively separating overlapping orders. The same 200mm lens becomes the camera lens as the final optical element focusing the spectrum into the detector. 

	\chapter{Software}
\label{ch:5}

\section{Wavelength Scale Model}

	High definition spectroscopy is partially possible due to the accurate identification of spectral lines in the image produced by the spectrograph. The true potential of the spectrograph can only be reached if the analysis can be done in a sub-pixel level.
	
\begin{figure}[ht]
	\begin{center}
	\includegraphics{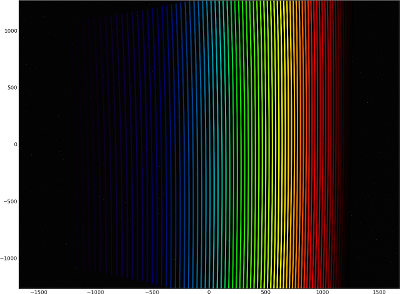}
	\end{center}
	\caption{A simulated flat emission source across the full spectrum of the Wavelength Scale Model. The parameters used correspond to the final fitting used on the solar spectrum.}
	\label{fig:Spectrum}
\end{figure}

	The main software component in charge of this task is the Wavelength Scale Model(WSM). It provides an accurate map of the detector chip that allows us to interpret the stellar spectra captured and the spectral lines it represents, see Figure~\ref{fig:Spectrum}. To that purpose, a forward model of the optical system was developed. The WSM replicates the distortions that the spectrograph produces on a beam of light from a given source. It traces the path of a monochromatic beam and computes its final location on the CCD detector. It is written in the freely available language \texttt{Python}. It allows the simulation of different sources by configuring the range of input wavelengths and the energy distribution.
	
	Alternative methods have been used to extract spectral information. DOECSLIT is a polynomial fitting of each order developed in IRAF(Image Reduction and Analysis Facility). It captures the location and extracts each order individually. Another option is to recreate the complete optical model using optical design software (i.e. Zemax). There is no direct way to fit individual spectral lines in this case. Scripting would be necessary adding the complexity and risking further complications like coordinate breaks. Finally, large scale projects use customized software developed in-house that is rarely freely available.

	The approach of the wavelength scale model for an Echelle grating is unique. It is neither an approximate relationship typical of Zemax models, which are applicable to design but not to the production spectrograph, and isn't a polynomial fit of wavelength versus pixel for each order, like DOECSLIT.

	The \texttt{main()} function takes 11 parameters that represent the degrees of freedom of the system. These values describe the orientation of the input vector and optical surfaces, as well as the physical properties of the grating and image distortion terms. 

	The CCD map simulated by a given source, is compared to known data captured by the spectrograph. A least square fitting procedure is performed to find the set of parameters that best fit the observations. This process gives us the right configuration necessary to interpret stellar observations and maximise the accuracy of the system. Spectral information can be accurately extracted as a consequence of a properly predicted wavelength identification.

\subsection{Reference Frame}
\label{sec:Ref_Frame}
\begin{figure}
\centering
\subfigure{
   \includegraphics[width=0.20\linewidth]{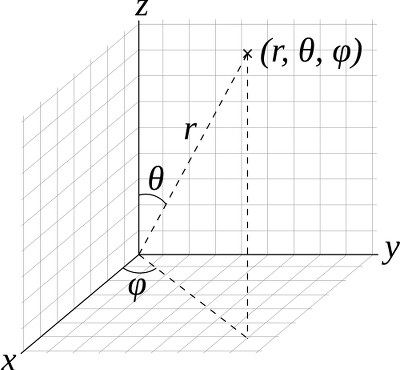}
   \label{fig:Axes}
 }
 \subfigure{
   \includegraphics[width=0.75\linewidth]{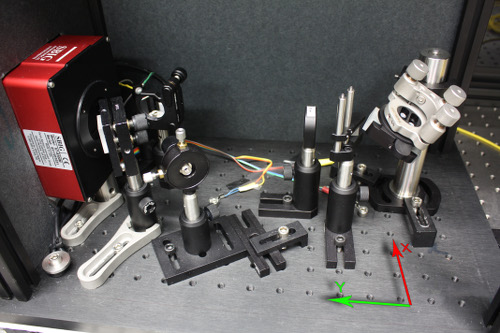}
   \label{fig:Spec_Axes}
 }
	\caption{Axes convention used to trace the beams through the spectrograph.}
	\label{fig:Axes_2}
\end{figure}

All calculations are based on a single reference frame, eliminating the potential error source of frame transformations. The axes are aligned with the camera lens-CCD axis. The x-axis runs along the width of the cage, with the positive side to the right when looking at the camera. The y-axis runs along the length of the cage, its positive direction towards the camera. The z-axis points upwards complying with a right hand convention, see Figure~\ref{fig:Axes_2}. 

The azimuth angle, $\phi$ has a range from 0 to 2$\pi$. It spans the x-y plane and has its 0 point in the positive x direction. It increases counter-clockwise as seen from the positive z-axis.

The polar angle, $\theta$ ranges from 0 to $\pi$, it's 0 point is in the positive z direction.

\subsection{The Input Parameters}

The orientation of the optical components is specified by unit vectors in Cartesian coordinates within the spectrograph's reference frame. 

The input beam describes the orientation of the radiation source after being collimated by the collimator/camera lens. Prism surfaces are described by their normal vectors. The grating requires 2 vectors to be fully characterised as the orientation of the grooves in space determine the portion of the input beam being affected. The blaze period is also provided. The last 2 parameters are the focal length of the system and the distortion parameter aimed to accurately plot the output beams on the CCD map. The units of the system are in microns and degrees except where otherwise stated.

\subsection{Snell's law in 3D}

When considering a 3-dimensional version of Snell's law, Equation~\ref{eq:Snell}, the definition of the plane of incidence becomes necessary. The 2-dimensional surface shared by the propagation vector and the normal of the boundary surface is the plane of incidence. The input vector is projected into this plane, restricting the problem to that described by Equation~\ref{eq:Snell} by losing a degree of freedom.

The code computes the change in angle of the beam in four steps. In the following steps $\theta_i$ is the incident angle, $\theta_r$ is the refraction angle, both measured  from the surface normal,  $\hat{u}$ is the input vector, $\hat{n}$ the surface normal, $\hat{p}$ the tangent vector to the surface in the plane of incidence and $\hat{v}$ is the output vector. 

First, $\hat{p}$ is found in explicit form

\begin{equation}
	\unit{p}=\frac{\unit{u}-\unit{n}(\unit{u} \cdot \unit{n})}{|\unit{u}-\unit{n}(\unit{u} \cdot \unit{n})|}.
\label{eq:p1}
\end{equation}  

Second, the incident angle is calculated using the fact that the dot product of input vector with the surface normal is the cosine of the subtended incident angle

\begin{equation}
	\theta_i=\arccos(\unit{u} \cdot \unit{n}).
\label{eq:theta}
\end{equation} 

Third, the refraction angle is calculated using Snell's law

\begin{equation}
	\theta_r=\arcsin(\sin \theta_i \frac{n_2}{n_1}).
\label{eq:theta'}
\end{equation} 

Fourth, the output vector is constructed by adding the normal and tangent vectors to the boundary surface multiplied by the cosine and sine of the refracted angle respectively

\begin{equation}
	\unit{v}= \unit{n} \cos \theta_r + \unit{p} \sin \theta_r.
\label{eq:v1}
\end{equation} 

The generalization of this process into the 3rd dimension adds flexibility and becomes particularly relevant in the second pass of the beam, once it has gained a significant vertical (z-direction) component from the grating.

\subsection{Grating Computation}

\subsubsection{Grating Orientation}

The orientation of the grooves in space is not uniquely defined by the vector normal to the grating surface. A second vector is used to remove this uncertainty. The vectors $\unit{s}$ and $\unit{l}$ are defined to run perpendicular and along the grooves respectively. The $\unit{s}$ vector is specified by it's polar and azimuthal angles, and an $\unit{l}$  specified by the angle it forms with the x-y.

\paragraph{The s vector}

From the provided polar and azimuthal angles, $\phi$ and $\theta$, the $\unit{s}$ vector can be constructed by a simple coordinate transformation

\begin{equation}
	\unit{s}=(\cos \phi \sin\theta , \sin\phi \sin\theta , \cos\theta)
	\nonumber
	\label{eq:s2}
\end{equation} 

\paragraph{The l vector}

To find the $\unit{l}$ vector 2 steps are needed. First we need to find a set of basis that span the plane perpendicular to $\unit{s}$. Second, define the orientation of $\unit{l}$ as a linear combination of these basis. 

The vectors $\unit{a}$ and $\unit{b}$ are introduced as auxiliary vectors. The derivation is found in Appendix \ref{App:gr_orientation}. 

The explicit from of $\unit{a}$ as a function of $\unit{s}$:

\begin{equation}
	\unit{a}=(\frac{s_y}{\sqrt{(s_x^2+s_y^2)}} ,-\frac{s_x}{\sqrt{(s_x^2+s_y^2)}}, 0)
	\nonumber
\end{equation} 

The vector $\unit{b}$ is simply the cross product between $\unit{a}$ and $\unit{s}$.

\begin{equation}
	\unit{b}=\unit{a} \times \unit{s}
	\nonumber
\end{equation} 

Having defined the basis to describe the $\unit{l}$ vector we find:
 
\begin{equation}
	\unit{l}=\cos\alpha \unit{a} + \sin\alpha \unit{b}
	\nonumber
\end{equation} 

where the angle $\alpha$ is one of the parameters of the system and it is measured from $\unit{a}$ to $\unit{b}$.


\subsubsection{The Grating Equation}

Knowing that $\unit{u}\cdot\unit{s}=\sin\theta_i$  and $\unit{v}\cdot\unit{s}=\sin\theta_r$, we can rewrite \ref{eq:Grating_Eq.} as

\begin{equation}
	\unit{u}\cdot\unit{s} - \unit{v}\cdot\unit{s} = \frac{n\lambda}{d}
\end{equation}

or

\begin{equation}
	\unit{v}\cdot\unit{s} = \unit{u}\cdot\unit{s} + \frac{n\lambda}{d}.
\label{eq:VectorGrating}
\end{equation} 
 
The analysis of the behaviour of a beam when it encounters a grating is divided in 2 main steps in the WSM. An initial step defines a unique position of the grating in the coordinate frame of the spectrograph, this is achieved by the creation of 2 auxiliary vectors that arise from 3 angles provided as parameters of the system. The second step is to compute the actual refraction pattern created by the grating. This last step is, in turn, divided in 2 steps by splitting the beam into components along the grooves and across them.

\subsubsection{Diffracted Beam}

Once the grating orientation is uniquely defined by its components across and perpendicular to the grooves, the problem of computing the grating equation can be divided into 2 parts corresponding to each of its components.

The effect of the grating in the direction parallel to the grooves is the same as a normal mirror. The angle subtended between the incident vector and the grooves, will be the same than the reflecting angle. 

\begin{equation}
	\unit{u} \cdot \unit{l} = \unit{v} \cdot \unit{l},
	\label{eq:l}
\end{equation} 

so the component in the $\unit{l}$ of the reflected beam will be 

\begin{equation}
	v_l = \unit{u}\cdot \unit{l}.
	\label{eq:vl}
\end{equation} 

The $\unit{s}$ component will depend on the order being computed. This is where the grating equation is finally needed. The $\unit{s}$ component reduces to 

\begin{equation}
	v_s = \unit{u}\cdot \unit{s} + \frac{n \lambda}{d}.
	\nonumber
\end{equation} 

Finally the $\unit{n}$ component is calculated by Pythagoras' theorem

\begin{equation}
	v_n = \sqrt{1-v_l^2-u_s^2}
	\nonumber
\end{equation} 

With the 3 components calculated, the output vector $\unit{v}$ can be constructed as
\begin{equation}
	\unit{v} = v_l \unit{l}+v_s \unit{s}+v_n (\unit{s}\times\unit{l})
	\label{eq:vn}
\end{equation}

\section{Other software modules}

\subsection{Spectrum Extraction}
\label{sec:SpecExtract}
Once the right parameters have been found, the extraction of the spectral information can be achieved one order at the time. The extraction function takes the output of the \texttt{main()} function, a 2-dimensional array containing a list in the format X-coordinate, Y-coordinate, wavelength, order. Initially, a single order is filtered. The second step is to create two interpolating functions that link the x and y coordinates, and the y coordinate and the wavelength. This allows us to track the order vertically following the deviations the it may have from a straight line. 
The next step is to loop through each of the Y-pixels to find it's corresponding X-value. The spectral pixel value is then calculated by integrating the width of the order. This has been measured to be $\sim$6 pixels in the solar spectrum presented in Section~\ref{sec:Spec}, but narrower width sizes have been found in the lab.
The resulting integrated flux as a function of Y is the plotted against the wavelength as a function of Y.

\subsection{Fitting}
\label{sec:fitting}

\begin{figure}[H]
	\begin{center}
		\includegraphics[width=1\linewidth]{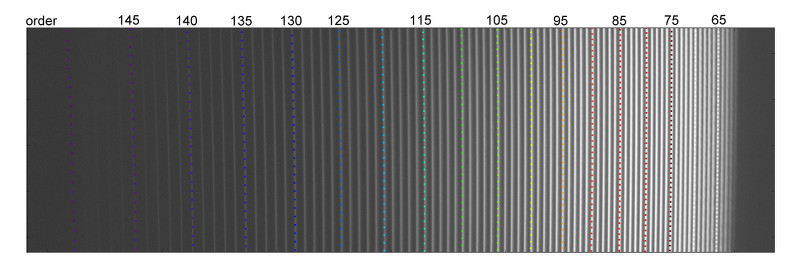}
	\end{center}
	\caption{The orders can be accurately identified once the right parameters have been found.}	
	\label{eq:orders}

\end{figure}

The initial estimation of the parameters that describe the optical system arises from measurements of the physical setup of the spectrograph. These values require a precise definition of a reference frame and a careful quantification of the degrees of freedom that the system will have. 
The orientation of the optical components and initial direction of the beam are described by nine angles. A pair of angles, corresponding to the polar and azimuthal inclination, are enough in each case to uniquely orientate the normal of the beam, first and second prism surfaces. The grating needs to be characterised by two angles, the normal would provide a correct orientation of the surface, but the orientation of the grooves would remain uncertain. The two vectors chosen to describe the orientation are perpendicular to the normal and oriented parallel and perpendicular to the grooves. Only three angles are necessary to describe the two vectors. One of them is described by its polar and azimuthal angles, and the second needs only its angular separation from a given reference plane, in this case the x-y plane, as it is perpendicular to the first vector. The last two parameters describe the groove period of the grating and the focal length of the system
In order to accurately calculate the location of a given wavelength will land on the CCD, the right parameters need to be found. This process can be approximated manually, but the final values need to be found by the fitting module, see Figure~\ref{eq:orders}. 
The first step of the fitting process is to capture spectra from a known source. Early in the project we realized that thorium argon was going to be too weak, and long exposures would be needed. The main calibrating source used for this project is mercury. The emission lines are captured and the location on the CCD sensor parametrized to be compared with the corresponding simulated version. The difference between physical and simulated results is the output of the \texttt{main\_errors()} function.
The fitting function, \texttt{doFit()}, loops over the \texttt{main\_errors()} function while changing the eleven input parameters. Finally, the output of the \texttt{doFit()} function is the vector that produces the closest results to the physical measurements becoming the fitted parameters.

\subsection{Image Calibration}
Preprocessing of the images before spectrum extraction is performed using the Image Calibration module. Several of the most common calibrating tasks are computed by this module including median, average, bias frame subtraction, dark frame scaling and subtraction and flat field calibration. The module is structured in functions that organize the information to be processed in each case. Most of the operations performed are simple operations between 2-dimensional arrays containing the information of the images and the function acts as a wrapper to the mathematical operation performed by an external package. 

\paragraph{Median and average}

For a given pixel across all images, the median or average pixel value is outputted to the final image.

\paragraph{Bias Frame}
The pixel count produced by a 0 second exposure represents the bias generated by the electronics. This frame is subtracted from the science frames to remove the count produced by this effect. 
\begin{equation}
	\text{Output\_Image}=\text{Science\_Image-Bias\_Frame}
\end{equation}

\paragraph{Dark Frame}
Time dependant charge is recorded in a Dark Frame. Using a bias subtracted dark frame allows us to resize the frame based on the exposure time. 
\begin{equation}
	\text{Output\_Image}=\text{Science\_Image -(Dark\_Frame}\times \text{Exposure\_Time)}
\end{equation}

\paragraph{Flat Field}
The uneven sensitivity of the pixels across CCD sensor is recorded by the flat field. 
\begin{equation}
	\text{Output\_Image}=\text{Science\_Image/Flat\_Frame}
\end{equation}

%
%
%
%
%
%
%
%
	\chapter{Results and Analysis}

\section{Fitting of the Wavelength Scale Model}
\label{sec:session1}

The eleven parameters of the WSM to uniquely characterise the system and the fitting procedure are described in \ref{sec:fitting}. The method presented here was used to extract the solar spectrum below. 

The mercury lines parametrized for calibration are:

\begin{table}[H]
  \centering
    \begin{tabular}{ccc}
    \toprule
    x-coordinate [pixel] & y-coordinate [pixel] & Wavelength [$\mu$m] \\
    \midrule
    -959.6 & 531.9 & 0.404656 \\
    -940.1 & -588  & 0.404656 \\
    -910.3 & 636.7 & 0.407783 \\
    -887.1 & -498.4 & 0.407783 \\
    -500  & 729.7 & 0.435833 \\
    -473.3 & -483.8 & 0.435833 \\
    467.4 & 203.5 & 0.546074 \\
    641.9 & -381.2 & 0.57696 \\
    599.3 & 1256.9 & 0.57696 \\
    641.4 & 189.4 & 0.579066 \\
    \bottomrule
    \end{tabular}%
     \caption{The 10 mercury lines used for calibration of the parameters corresponding to the session of solar spectrum acquired for Section~\ref{sec:Spec}.}
  \label{tab:HG_lines_param}%
\end{table}%

Using the the information on Table~\ref{tab:HG_lines_param}, the fitting module produced the following vector
\\\\
\texttt{p = [272.31422902, 90.7157937, 59.6543365, 90.21334551, 89.67646101, 89.82098015, 68.0936684,  65.33694031, 1.19265536, 31.50321471, 199.13548823]}
\\\\
that represents the eleven parameters that characterise the configuration of the spectrograph at the time of the session. The physical meaning of the vector is presented in Table~\ref{tab:P}. The coordinate system used is described in Section~\ref{sec:Ref_Frame} and the $\alpha$ angle in Appendix~\ref{App:gr_orientation}. 

\begin{table}[H]
  \centering

    \begin{tabular}{c}
    \toprule
    Injected Beam               \\
	$\phi$=272.31$^{\circ}$ \\
    $\theta$ = 90.76$^{\circ}$ \\
    \midrule
    Prism (Face 1)                \\
          $\phi$   =59.65$^{\circ}$ \\
           $\theta$ =90.21$^{\circ}$ \\
    \midrule
    Prism (Face 2)               \\
           $\phi$   = 89.68$^{\circ}$ \\
           $\theta$ = 89.82$^{\circ}$ \\
    \midrule
    Grating                \\
           $\phi$   = 68.09$^{\circ}$ \\
           $\theta$ = 65.34$^{\circ}$ \\
          $\alpha$ = 1.19$^{\circ}$ \\
    \midrule
   		   Blaze Period= 31.50$\mu$m \\
    \midrule
    	   Focal Length = 199.16mm \\
    \bottomrule
    \end{tabular}%
      \caption{The best fitted parameters allow us to predict accurately the location of the different spectral lines.}
  \label{tab:P}%
\end{table}%

\begin{figure}[H]
\centering
\subfigure[The 0.404nm and 0.407nm lines of mercury.]{
   \includegraphics[width=0.25\linewidth]{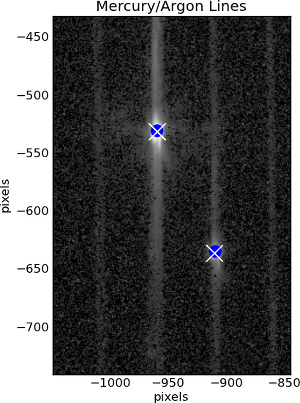}
 }
 \subfigure[The 0.546nm and 0.579nm lines of mercury.]{
   \includegraphics[width=0.35\linewidth]{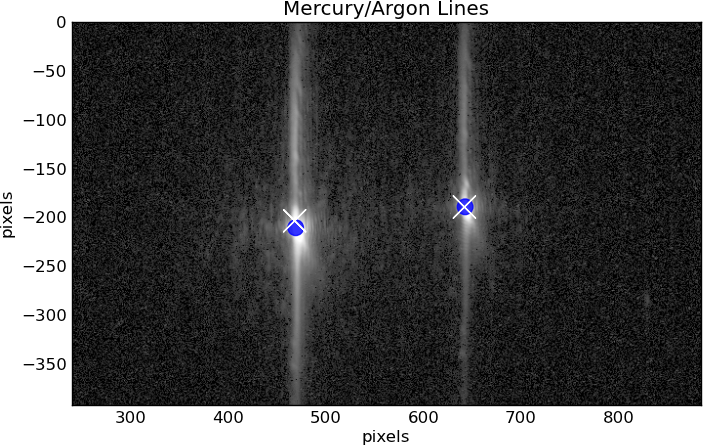}
 }
 	\caption{Mercury spectral lines and their location predicted by the Wavelength Scale Model. Blue dots are calculated locations.}
	\label{fig:Hg_fit}
\end{figure}

Plotting the parametrized coordinates with the calculated lines in Figure~\ref{fig:Hg_fit}, allows us to see the fitting accuracy.

\begin{figure}[H]
\centering
   \includegraphics[width=0.45\linewidth]{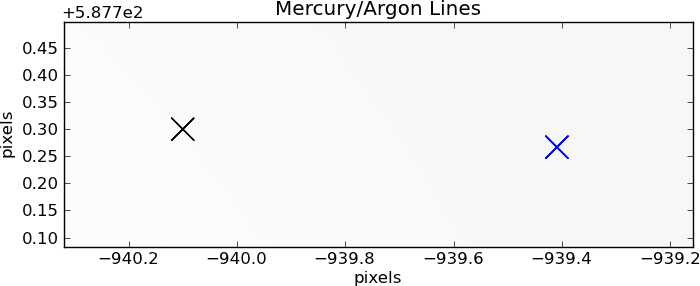}
	\caption{The parametrized and calculated location of a single emission line showing the error in the fitting.}
	\label{fig:404_fit}
\end{figure}

An example of the error in the fitting model is shown in Figure~\ref{fig:404_fit}. The predicted value is shown in blue, and the parametrized value in black. The error in the y direction is $\sim$200$^{th}$ of a pixel and in the x direction is $\sim$~1.2 pixels.

\begin{figure}[H]
\centering
   \includegraphics[width=0.6\linewidth]{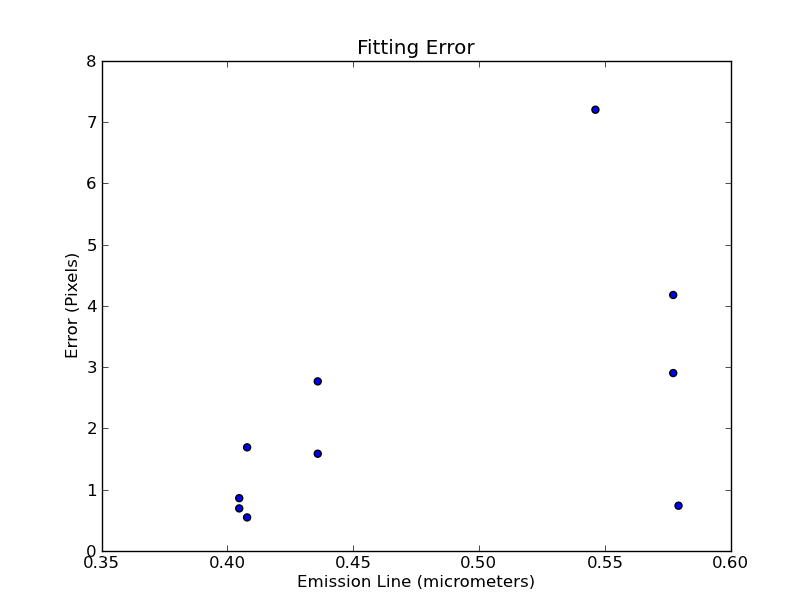}
	\caption{The pixel distance between the predicted value and the parametrized value for the ten emission lines used in this model.}
	\label{fig:Error1}
\end{figure}

The ten pixels parametrized in this model are shown in Figure~\ref{fig:Error1}.

\begin{figure}[H]
\centering
   \includegraphics[width=1\linewidth]{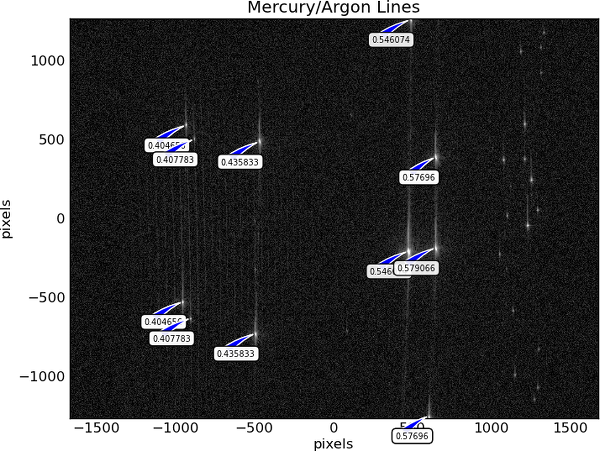}
	\caption{The strongest mercury lines are identified in the RHEA spectrograph. The label overlay is automatically produced by the software in the predicted points of the image, based on the eleven parameter physical spectrograph model.}
	\label{fig:Mercury}
\end{figure}

Once the right parameters are found, all lines can be identified, see Figure~\ref{fig:Mercury}.

\section{Spectrum Extraction}
\label{sec:Spec}
Using the parameters presented in Section~\ref{sec:session1}, the physical wavelength model can identify the wavelength and order corresponding to any pixel value. By doing this, we can extract and compile the full spectrum of any source projected through the spectrograph while the calibration parameters are still valid.  

\subsection{Mercury Spectrum}

This analysis corresponds to the same session than the solar spectrum presented below, however, the image produced is not according to the standards measured in the lab. It was noticed after the measurements that the resolution achieved is well below the estimated and also below the best attained. Based on previous experiments, the movement of the spectrograph, thermal changes and optical misalignments, all play a role in the final quality obtained.  Nonetheless, the fitting parameters have been found successfully, see Figure~\ref{fig:Mercury}, as the solar spectrum in the previous section shows.

\begin{figure}[H]
\centering
\subfigure{
   \includegraphics[width=0.45\linewidth]{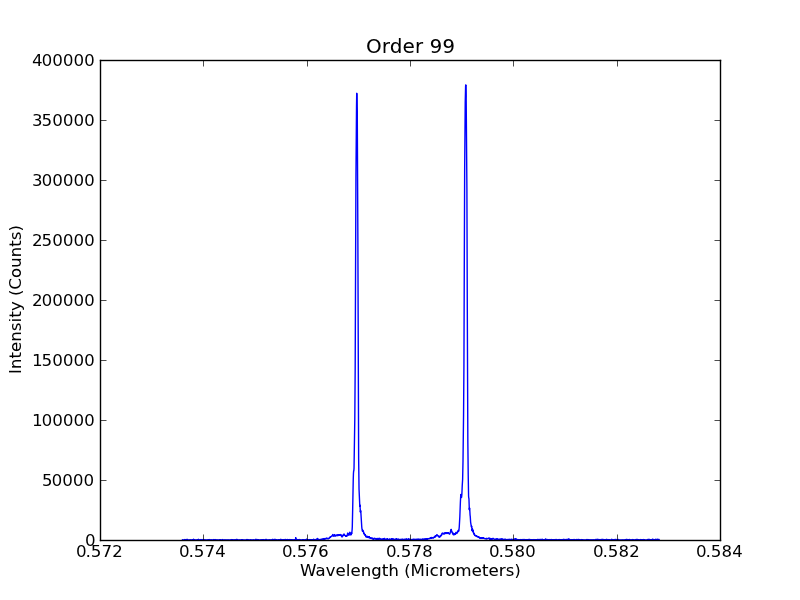}
 }
 \subfigure{
   \includegraphics[width=0.45\linewidth]{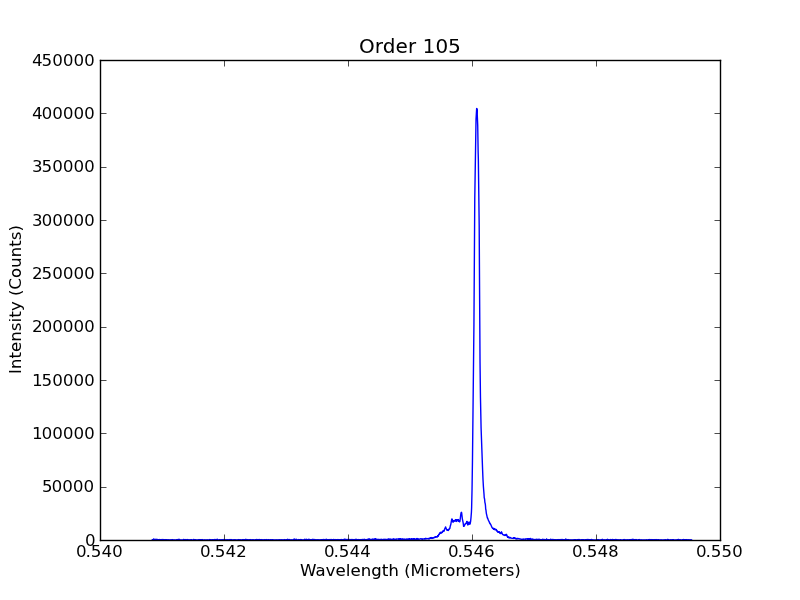}
 }
 \subfigure{
   \includegraphics[width=0.45\linewidth]{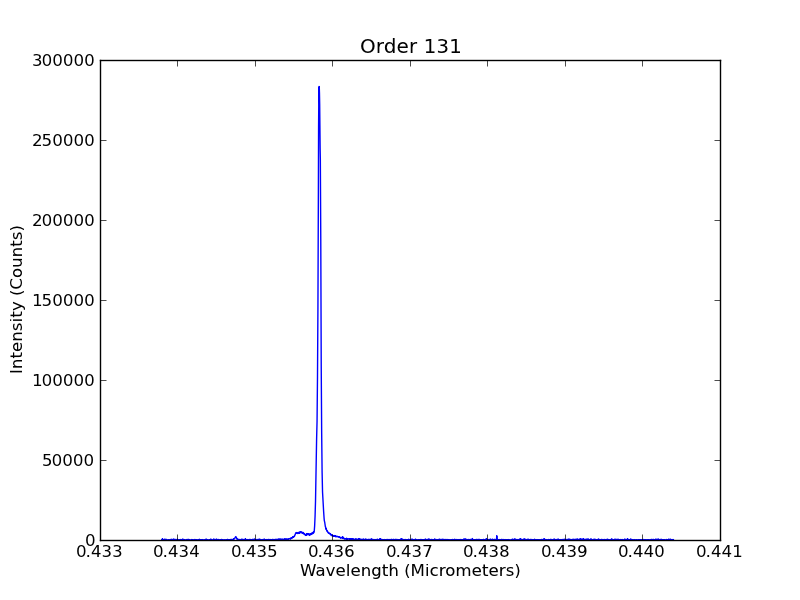}
 }
 \subfigure{
   \includegraphics[width=0.45\linewidth]{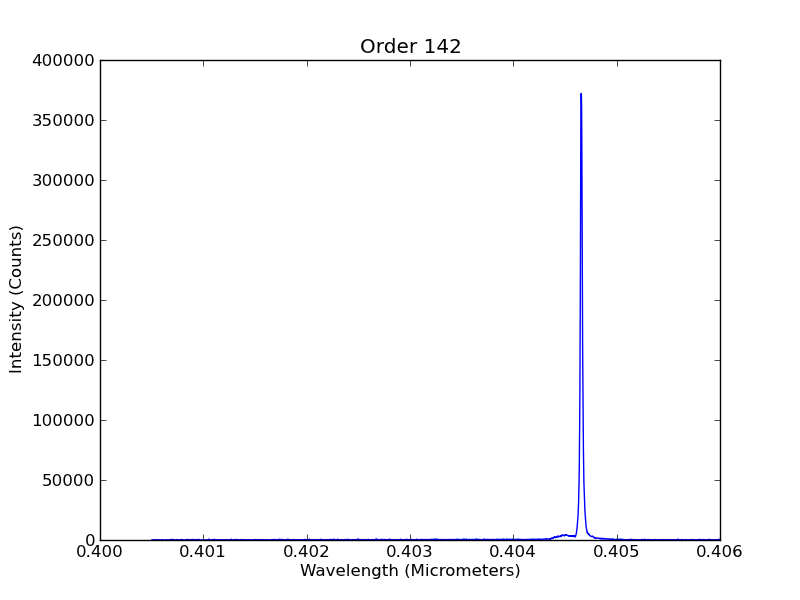}
 }
	\caption{Mercury light presents very localized, distributed emission lines making it an ideal calibrating tool. Order 105 shows the well known 546.074 nanometer line.}
	\label{fig:Hg_Orders}
\end{figure}

\begin{figure}[H]
\centering
   \includegraphics[width=1\linewidth]{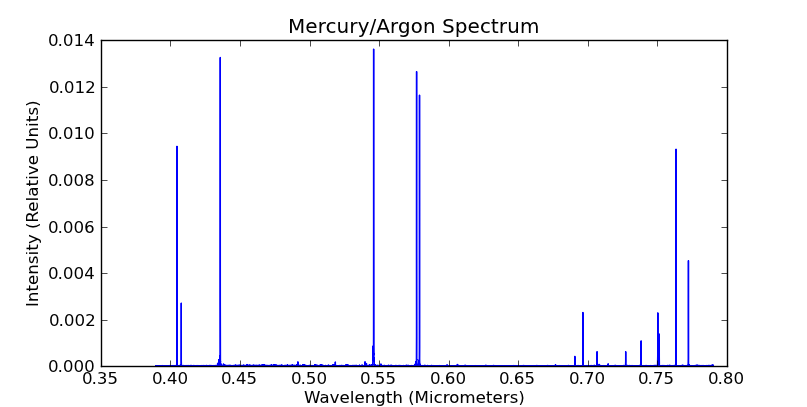}
	\caption{The complete integration of the mercury spectrum across the wavelength range attainable by the RHEA spectrograph.}
	\label{fig:Full_Mercury}
\end{figure}

Figures~\ref{fig:Hg_Orders}~and~\ref{fig:Full_Mercury} show the main emission lines of mercury. In the long wavelength part of FIgure~\ref{fig:Full_Mercury} some argon lines can be noticed.

\subsection{Solar Spectrum}

\begin{figure}[H]
\centering
   \includegraphics[width=.5\linewidth]{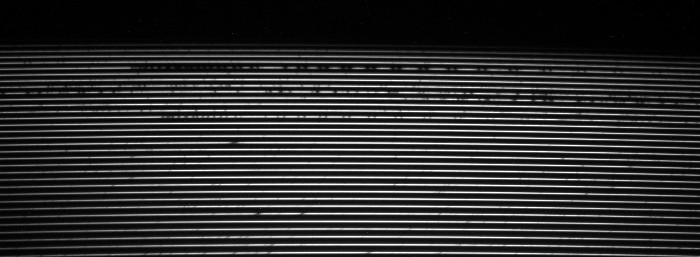}
	\caption{A section of the calibrated solar spectrum from a combination of 3 exposures of 30 minutes each (rotated).}
	\label{fig:SS}
\end{figure}

The sky scatters solar radiation making it a good target for capturing the sun's spectrum and characterising the spectrograph. It is representative of a well-aligned and focused star of visual magnitude 3. It allows us to focus on spectrograph performance without the extra varables added by the other components(i.e. Telescope and Fibre Feed). The solar spectrum presented here was captured through a 3 meter long, 9.6 $\mu$m core fibre exposed to the open sky. A 30 minute calibrated image presents recognizable Fraunhofer lines, see Figure~\ref{fig:SS}.

\begin{figure}[H]
\centering
   \includegraphics[width=1\linewidth]{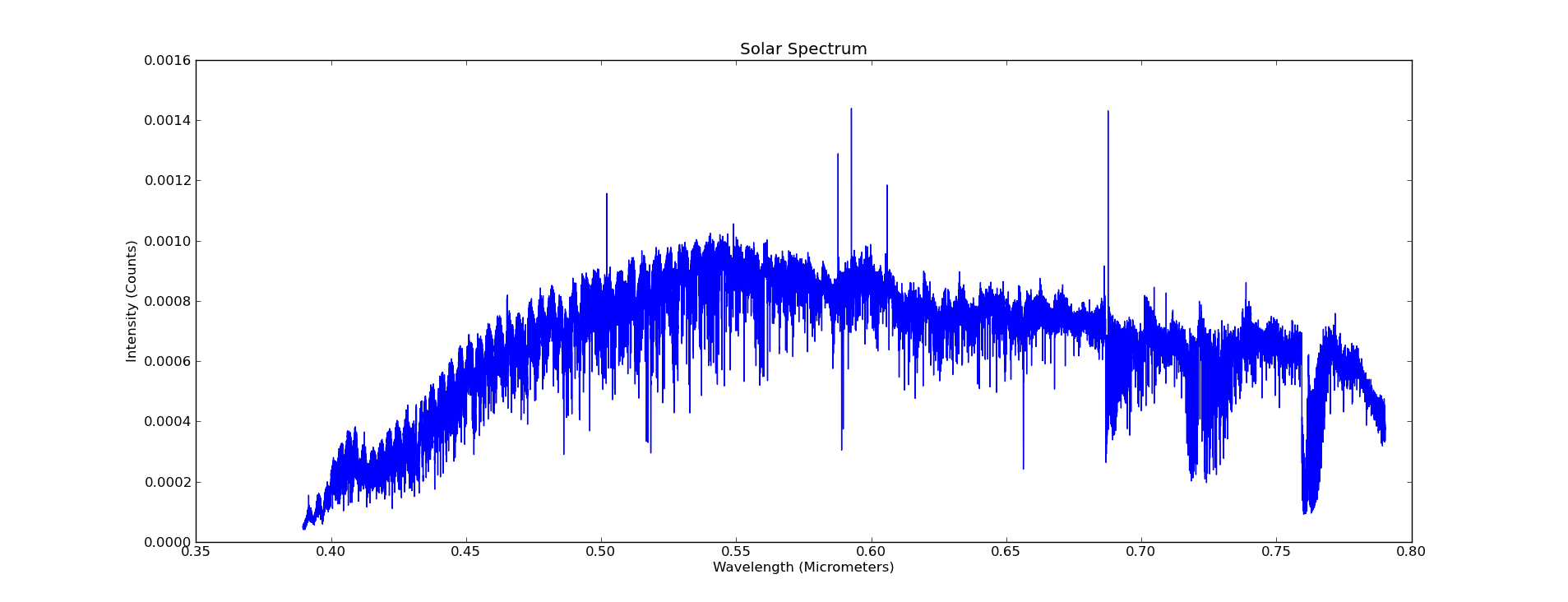}
	\caption{The complete solar spectrum integrated over all orders. The flat calibration is corrected using a black body curve at 3000 Kelvin.}
	\label{fig:Full_Solar_Spectrum}
\end{figure}

\begin{figure}[H]
\centering
   \includegraphics[width=1\linewidth]{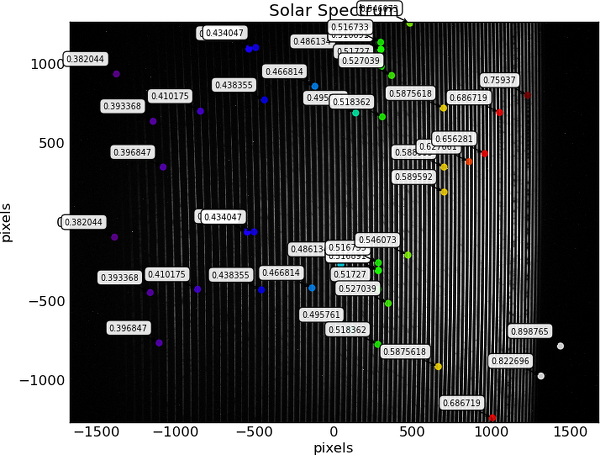}
	\caption{The most prominent Fraunhofer lines over the calibrated solar spectrum.}
	\label{fig:Fraunhofer}
\end{figure}

The overlay of the most prominent lines over the captured spectrum shows their location on the CCD, see Figure~\ref{fig:Fraunhofer}.

The different orders are individually extracted by the method specified in Chapter~\ref{sec:SpecExtract}. This method can produce high resolution spectrum, see Figure~\ref{fig:NA_Doublet}. Some of the orders presenting the most prominent features are presented below, see Figure~\ref{fig:Stelar_Orders}.

\begin{figure}[H]
\centering
\subfigure[Large oxygen absorption region including atmospheric absorption beyond the 0.762 $\mu$m wavelength.]{
   \includegraphics[width=0.45\linewidth]{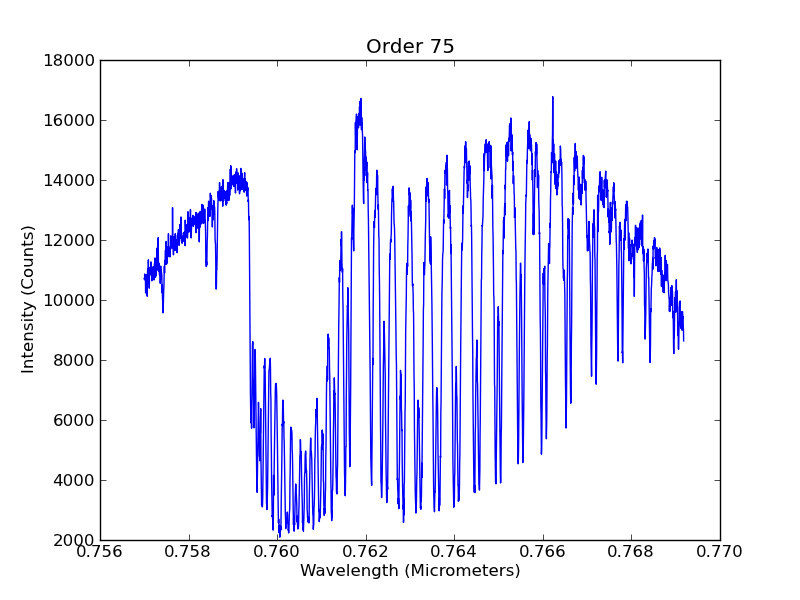}
 }
 \subfigure[H-alpha absorption line from hydrogen atoms as a consequence of the electron orbit decay from the 3$^{rd}$ to the 2$^{nd}$ energy level.]{
   \includegraphics[width=0.45\linewidth]{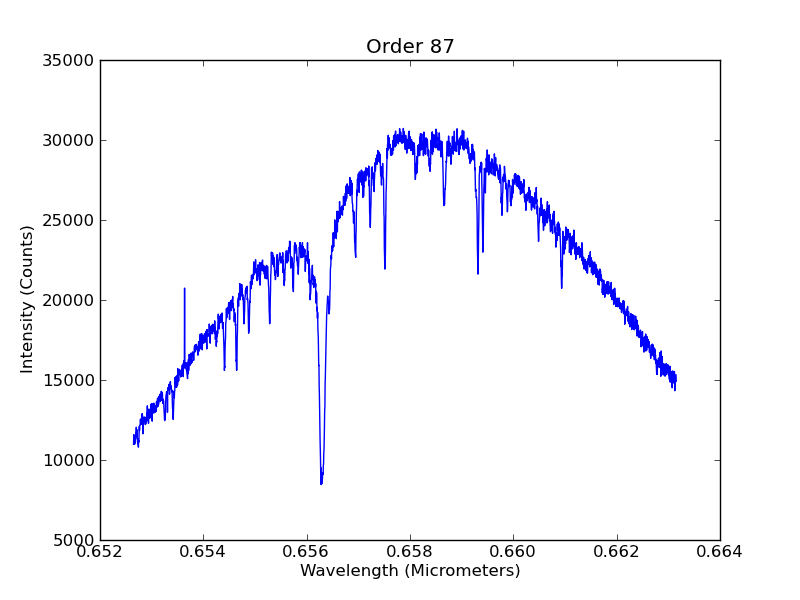}
 }
 \subfigure[The sodium doublet arises from the small energy difference released when electrons with different angular momentum transfer from the 3p to the 3s orbit levels.]{
   \includegraphics[width=0.45\linewidth]{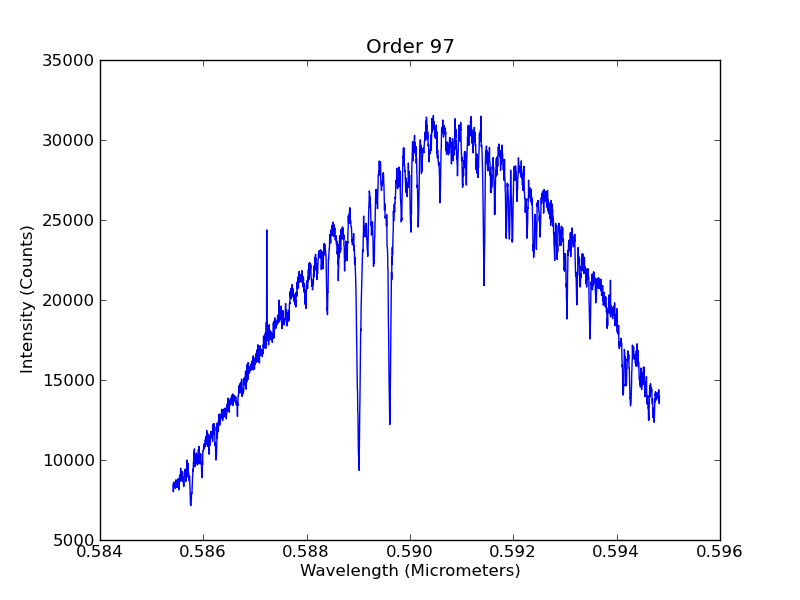}
 }
 \subfigure[The wide iron absorption line is shown at .527$\mu$m wavelength.]{
   \includegraphics[width=0.45\linewidth]{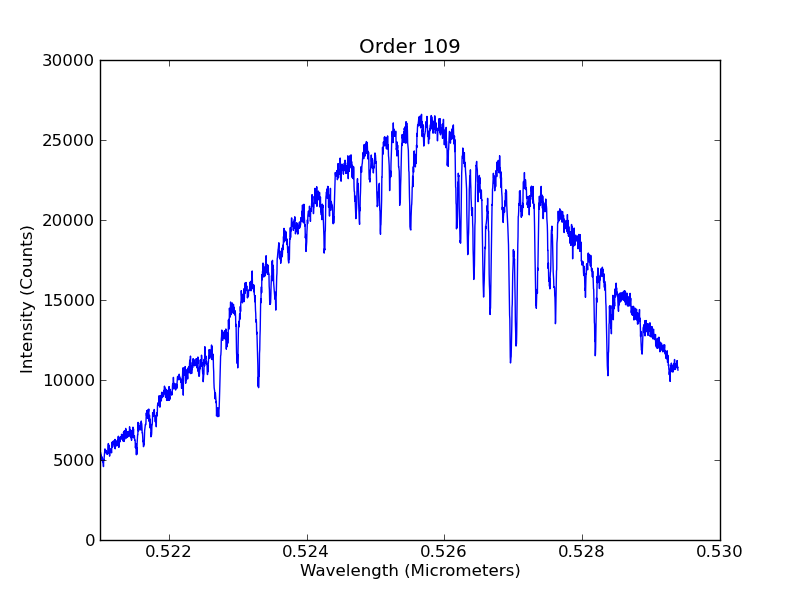}
 }
	\caption{Some of the most significant absorption lines detected in the solar spectrum.}
	\label{fig:Stelar_Orders}
\end{figure}

Figure \ref{fig:Full_Solar_Spectrum} shows the integrated compilation of all orders produced by the RHEA spectrograph. Each individual order has been calibrated by a flat frame created by injecting a tungsten light through he spectrograph. Each flat order was individually collected and subtracted from the corresponding order of the sky spectrum. Finally all orders where compiled and cleaned of overlapping regions in Figure~\ref{fig:Full_Solar_Spectrum}. 

\begin{figure}[H]
\centering
   \includegraphics[width=.7\linewidth]{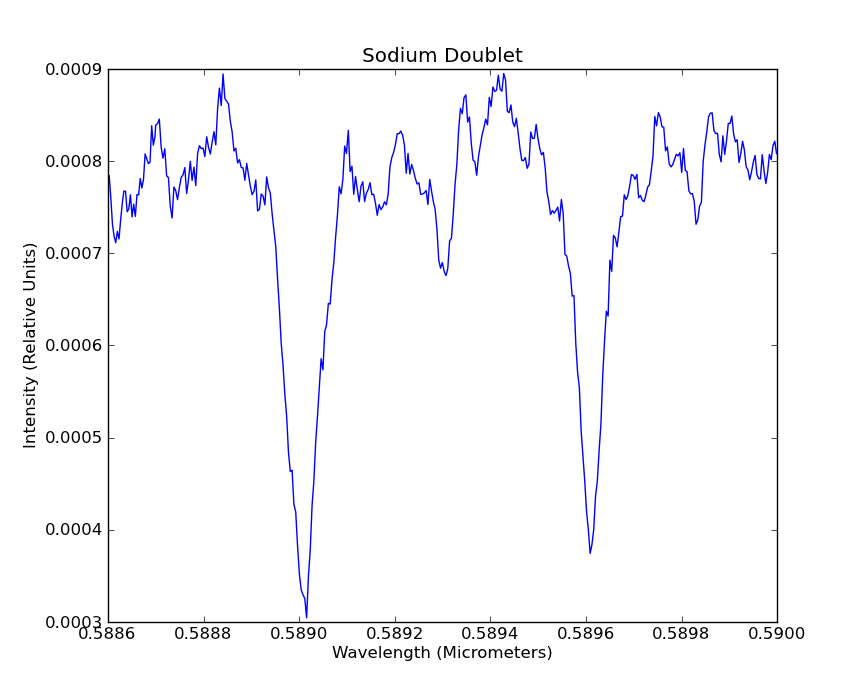}
	\caption{A calibrated high resolution image of the sodium doublet. The FWHM of each absorption line is of the order of a tenth of a nanometer.}
	\label{fig:NA_Doublet}
\end{figure}

\subsection{Arcturus Spectrum}

A single observing session could be used to observe the spectrum of Arcturus before the spectrograph's camera was moved due to a bump. The tracking of the telescope and the alignment of the fibre feed allowed only for short exposures. The image used to extract this spectrum was calibrated by averaging three exposures of one minute and subtracting a one minute dark frame. The spectral pixel count only reaches 1600, which is equivalent to

\begin{equation}
	\Delta\lambda = \frac{\lambda}{R}=\frac{0.656\mu\text{m}}{145000}=4.5 \times 10^{-6} \mu\text{m},
\end{equation}

\begin{equation} 
	F_{Arc}= \frac{\text{Flux} \times 0.38 \text{ph}}{\Delta\lambda\times60 \text{s}}= 2.25185 \times 10^6 \text{ph/}\mu \text{m/s}
\end{equation}

\begin{figure}[H]
\centering
   \includegraphics[width=.7\linewidth]{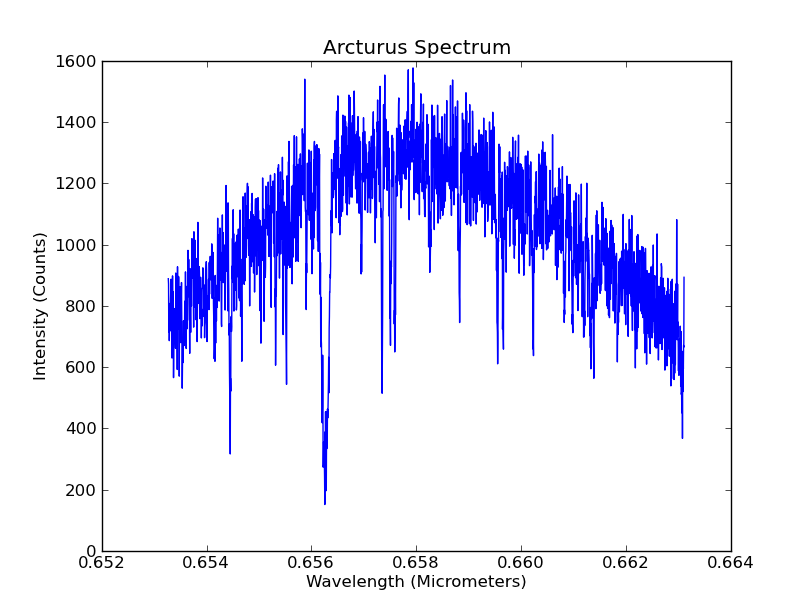}
	\caption{Order 87 of Arcturus spectrum. The H-alpha absorption line can be noticed at 0.656 micrometers.}
	\label{fig:Arcturus}
\end{figure}

\section{Spectrograph Throughput}

A key measurement of the spectrograph's performance is the throughput. The ratio of the amount of light entering the spectrograph to the amount of light received at the CCD sensor is of great importance in astronomy in general. Particularly in the case of single mode injected spectrograph, this importance is increased by the limitations imposed by the small core of the feeding fibre. The overall throughput of a system is wavelength dependent as it is a consequence of the dependency of the optical components. 

\subsection{Estimated}
\begin{figure}[H]
\centering
   \includegraphics[width=1\linewidth]{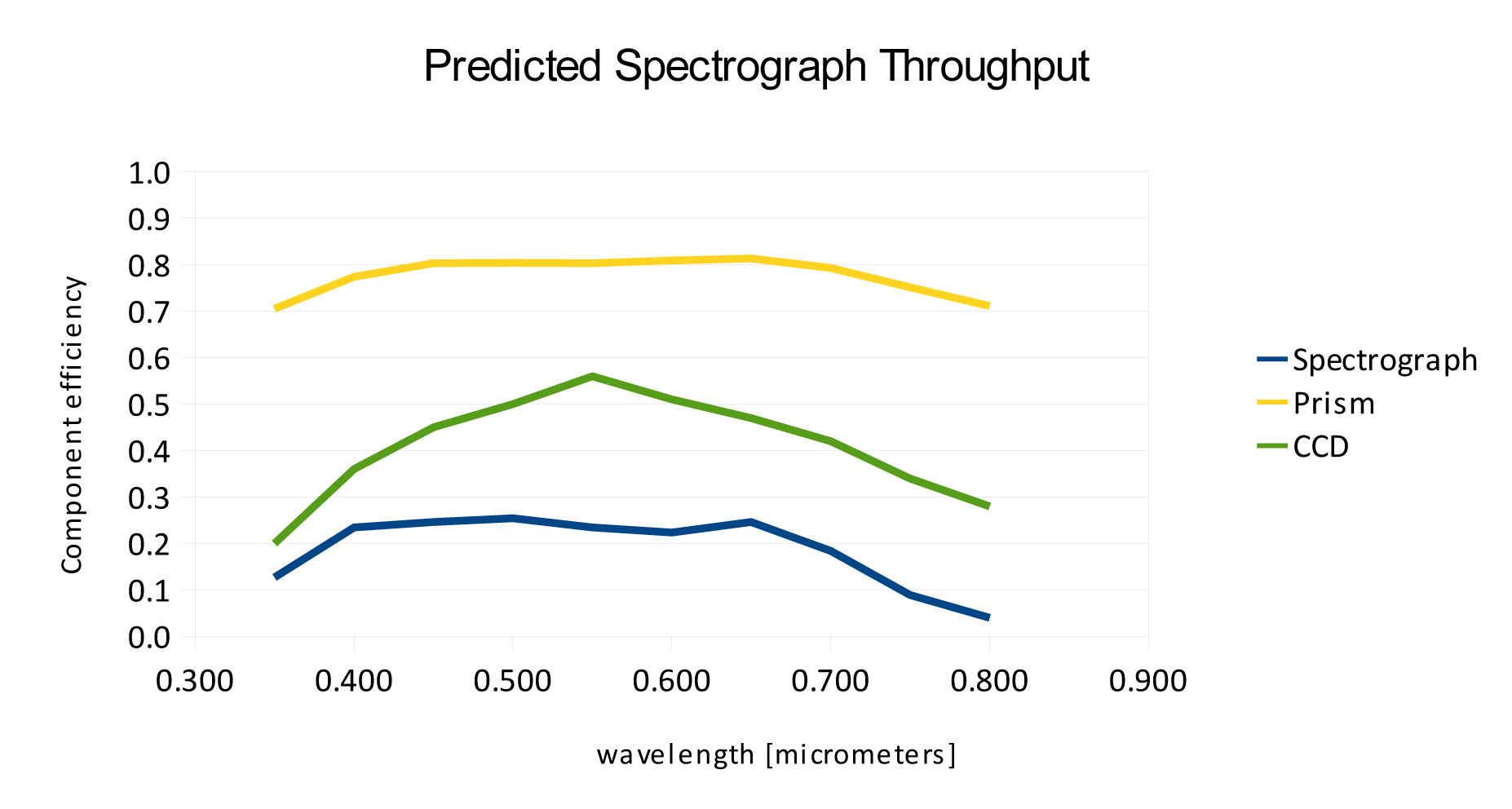}
	\caption{The predicted wavelength dependant throughput of the systems shows its maximum efficiency in in the range between 0.42 $\mu$m and 0.65 $\mu$m.}
	\label{fig:Throughput}
\end{figure}

\subsection{Measurements}

In order to calculate the throughput of the system experimentally, the flux collected by the fibre optic. 

For the throughput calculation, the following values were assumed :

\begin{equation}
	\begin{array}{rcl}
		Mag_ {\odot} &=& -26.75 \\
		F_0(at 0.55\mu m) &=&  1.08\times 10^{11} \text{ph/m}^2/\mu \text {m/s} \\
		S_{0m} &=&  0.243 \text{mag/airmass}\approx 20 \% \\
		NA &=& 0.13
	\end{array}
\end{equation}

where $Mag_ {\odot}$ is the relative magnitude of the Sun as perceived from earth \citep{cox_allens_1999}, $F_0$ is the flux of a 0th magnitude star at 550 nanometers, S$_{0m}$ is the scattering of blue sky at sea level \citep{rufener_evolution_1986} and NA is the numerical aperture of the optical fibre used to measure the flux. 

The flux of the Sun as it reaches the top of the Earth's atmosphere, $F_{\odot}$, is

\begin{equation}
	F_{\odot} = F_0 \times 10^{-0.4 \times Mag_{\odot}} = 5.1692 \times 10^{21} \text{ph/m}^2/\mu \text{m/s}.
\end{equation}

The portion that reaches the surface, F$_{\odot(surf)}$, is

\begin{equation}
	F_{\odot(surf)} = F_{\odot} \times 0.2 = 1.03384 \times 10^{21} \text{ph/m}^2/\mu \text {m/s}.
\end{equation}

The flux at the entrance of the single mode fibre(460HP), F$_{460HP}$, having a mode field diameter (MFD) of 3.5$\mu$m is 

\begin{equation}
	F_{460HP} = F_{\odot(surf)} \frac{(NA^2 \pi)(\pi\times (\nicefrac{\text{MFD}}{2})^2)}{4\pi}        = 4.20249 \times 10^7 \text{ph/}\mu \text{m/s}.
\end{equation}

The spectral pixel bandwidth of the REAH spectrograph at 0.55$\mu$m is
 
\begin{equation}
	\Delta\lambda = \frac{\lambda}{R}=\frac{0.55}{145000}=3.8 \times 10^{-6} \mu\text{m},
\end{equation}

so for a 30 minute exposure and a CCD gain value of 0.38, the expected count at 0.55$\mu$m is

\begin{equation}
	\text{Flux} =\frac{F_{460HP} \Delta\lambda \times 1800}{0.38}= 755075 
\end{equation}

The measurements showed a count of F$_{550}\approx$27000, so the throughput at 0.55$\mu$m, T$_{550}$, is 

\begin{equation}
	T_{550} =\frac{F_{550}}{Flux}= 0.0397
\end{equation}

This is well below the estimated value of $\approx$0.23.

\section{Spectral Resolution}

The spectral resolution of a spectrograph is the capacity to identify two neighbouring wavelengths. It is a measurement of how close these wavelenghts can be and still be identified as individuals. 

\subsection{Estimated}

Using the small angle approximation, $\theta \approx \sin\theta$, the angle projected into a single pixel in the focal plane of the system given by

\begin{equation}
	\Delta \theta_{\rm{pix}}\approx \frac{\Delta x_{\rm pix}}{f}
\end{equation}

where $f$ is the focal length of the camera. So

\begin{equation}
	\Delta \theta_{\rm{pix}}=\frac{\SI{5.4}{\micro\metre} }{200\text{mm}}=2.7\times10^{-5} 
\end{equation}

From \ref{eq:dtheta}

\begin{equation}
	R_{\rm pix}=\frac{2\tan{\theta}}{\Delta\theta_r}
\end{equation}

setting $\theta=63\deg$, the blaze angle, yields

\begin{equation}
	R_{\rm pix} = 145379.
	\label{eq:r_pix}
\end{equation}

Working at diffraction limit, the FWHM of an airy disk is given by $\frac{\lambda}{D}$. Using the 546nm emission line from a mercury lamp as a reference, the expected value in a detector with \SI{5.4}{\micro\metre} pixels is

\begin{equation}
	{\rm FWHM}_{\rm pix} = \frac{\lambda f}{D \Delta x_{\rm pix}}=\frac{546\rm nm 200 \rm mm}{9\text{mm}\SI{5.4}{\micro\metre}}\approx2.02\,\text{px}.
	\label{eq:Res}
\end{equation}

The measured value is expected to be larger than that, see comments in section~\ref{sec:Measurements}, so for a FWHM that spans over 3 pixels this means 

\begin{equation}
	R \approx 50000.
\end{equation}

\subsection{Measurements}
\label{sec:Measurements}

The spectral resolution of the system is computed for different spectral lines by fitting a Gaussian function to each line. The plot of each fitting is presented with the results.

\begin{figure}[H]
\centering
  \subfigure{
   \includegraphics[width=0.45\linewidth]{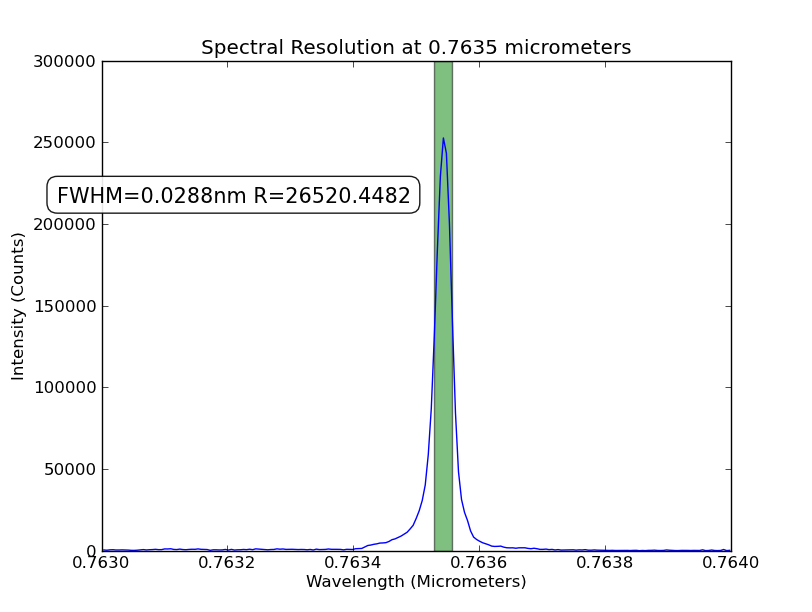}
 }
 \subfigure{
   \includegraphics[width=0.45\linewidth]{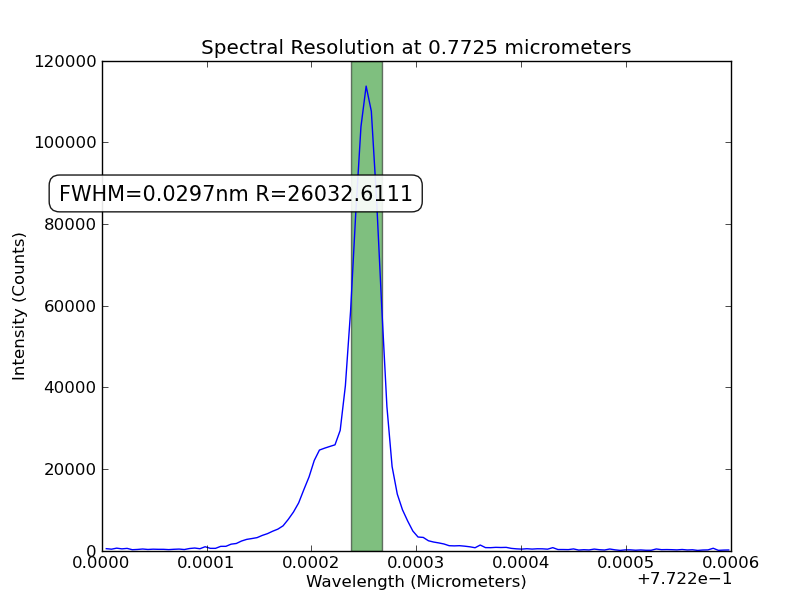}
 }
 \caption{The spectral resolution of the RHEA spectrograph is calculated by fitting a Gaussian function through the spectrum of an emission line and measuring its FWHM.}
 \label{fig:Hg_Res}
\end{figure}

Several mercury lines across the spectrum are measured to establish the spectral resolution of the spectrograph. The narrow lines are an ideal source to caracterized the spectrograph's response in different regions of the spectrum. The resolution was found to be stable across the wavelength range. This was unexpected and it is a consequence of a degrading image towards the short wavelength region. 

\begin{figure}[H]
\centering
   \includegraphics[width=1\linewidth]{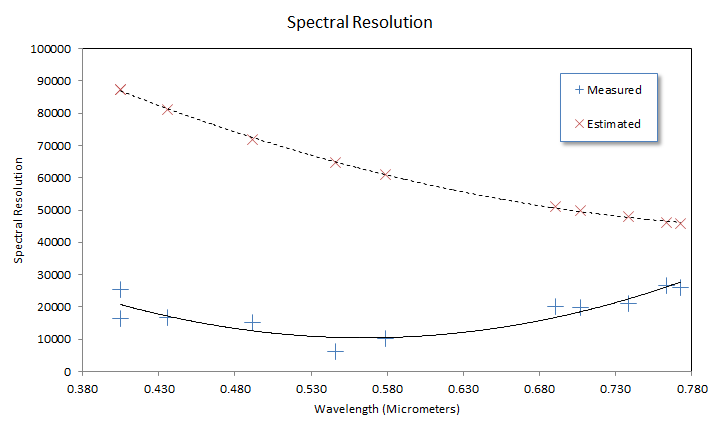}
	\caption{A comparison between the expected and achieved spectral resolutions. Measured values arise from the automatic Gaussian fitting of the main spectral lines of mercury and argon, expected resolution is calculated from taking the ratio of Equation~\ref{eq:r_pix} and Equation~\ref{eq:Res}. Image aberration was noticed in the blue end of the spectrum and only the argon emission lines approximate the expected resolution.}
	\label{fig:Spec_Res}
\end{figure}

From the results obtained across the spectrum, a decrease in the spectral resolution towards the shorter wavelengths is a consequence of a misalignment of the system. All mercury emission lines were blurred. It was only the weak argon emission lines at the longest wavelengths that the approached the expected resolution.

\begin{figure}[H]
\centering
  \subfigure[Transversal profile of the emission line in the widened direction.]{
   \includegraphics[width=0.45\linewidth]{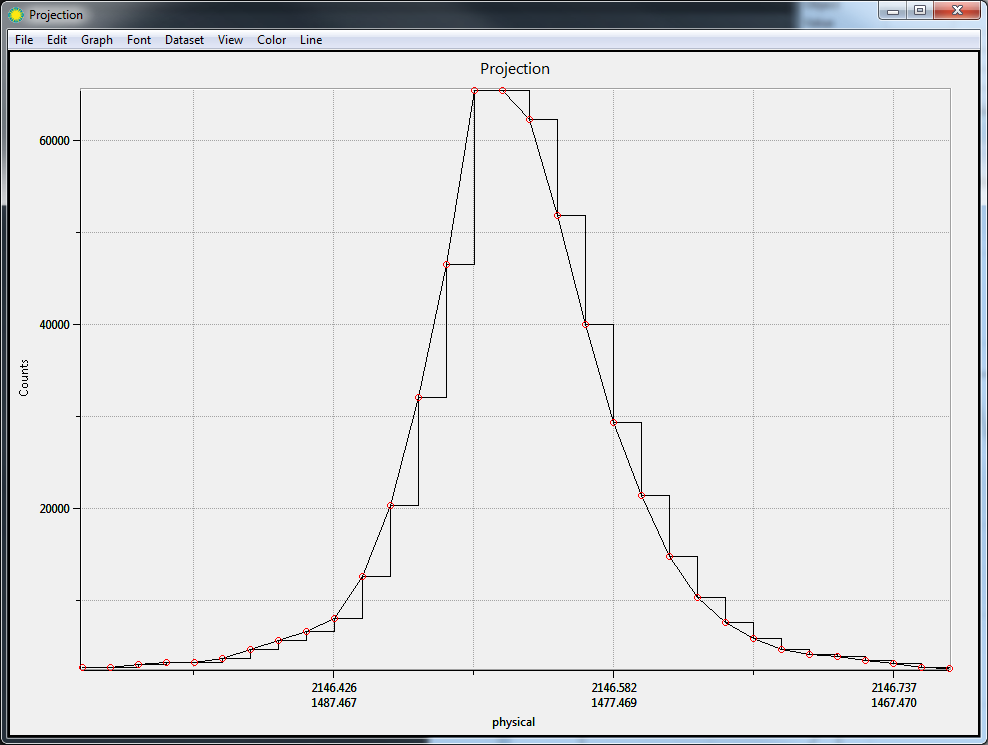}
 }
 \subfigure[Transversal profile of the emission line showing a width according to the expected resolution.]{
   \includegraphics[width=0.45\linewidth]{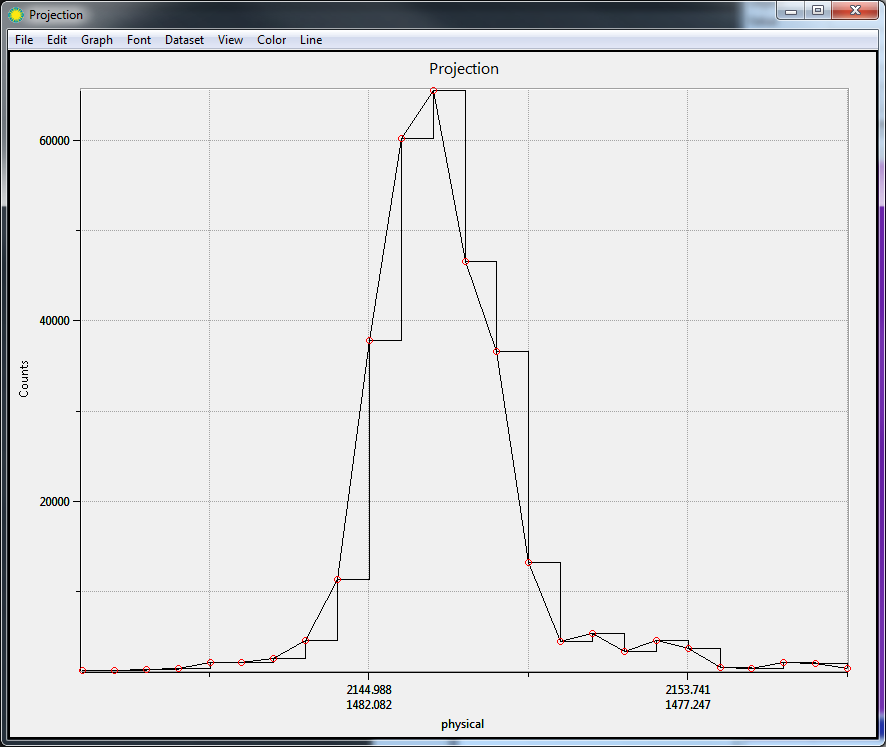}
 }
 \subfigure[Emission line showing a non uniform aberration.]{
   \includegraphics[width=0.45\linewidth]{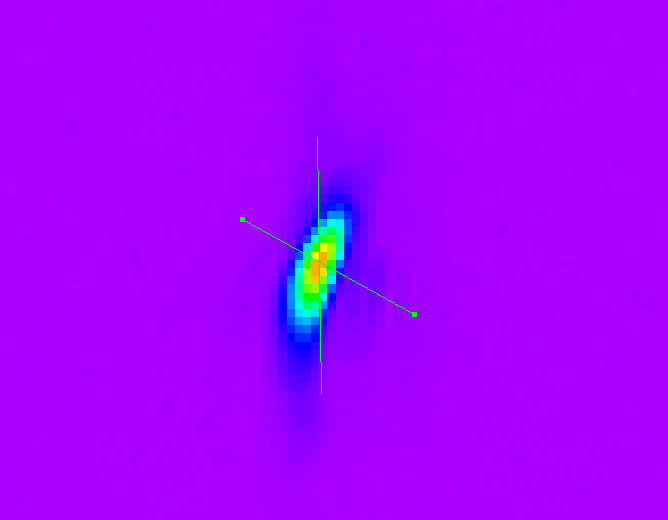}
 }
	\caption{Two different emission lines from the same exposure and at the same scale. The aberration in the 546 nanometer line on the left is clearly noticeable.}
 \label{fig:Hg_Compare}
\end{figure}

The orientation of the image is oblique with respect of the direction of the orders. This prevents us from reaching the expected resolution. A comparison of the image profile at different angles shows that in the narrowest direction the FWHM approaches the $\sim$3px size that the spectral pixel was expected to measure.

\section{Thermal Stability}

\begin{figure}[H]
\centering
   \includegraphics[width=.8\linewidth]{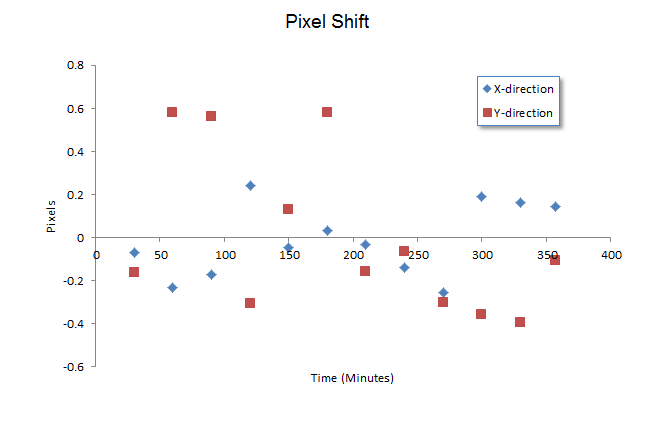}
	\caption{Pixel shift measured over 360 minutes.}
	\label{fig:Pixel}
\end{figure}

In order to test the thermal response of the spectrograph, 720 images over a period of 360 minutes were acquired. Mercury lines were recorded en each case. The position of the centroid of the 546 nanometer line was compared across images, see Figure~\ref{fig:Pixel}.

Despite exhibiting large scatter in the data from both axes of the original image, the shift in the Y axis is noticeable larger. This axis is  along a given order and a shift in this direction could imply a shift in the wavelength of the source. The mercury lamp used has a 30 minute stabilization time. This could account for the changes at the beginning of the measurements. Nonetheless, similar shifts occur later rendering the measurements inconclusive.

	\chapter{Conclusions}

The search for exoplanets is a fast growing field in astronomy that is producing quantitative results at unprecedented rates. Nonetheless, the range of planetary configurations that is being studied is biased by instrument limitations and operational costs. The RHEA spectrograph is a large project that includes several components beyond the scope of this thesis. The ultimate goal is to develop a replicable spectrograph with enough sensitivity to search for extrasolar planets with off-the-shelf components using 0.4 to 1 meter class telescopes. This concept could be developed to form a network spread across the world feeding a centralized data centre, effectively filling-in a gap in the range of systems studied.

This honours thesis focused on quantifying and testing the operational capacity of the first working prototype of the spectrograph. This was achieved by the development of a physical wavelength software model that could simulate the behaviour of the spectrograph to accurately extract the produced spectra.

The spectrograph prototype required a considerable amount of development, as the original design was changed at the beginning of the project to a more compact configuration. Several technical difficulties were presented that required realignment. The physical integrity of the spectrograph while being transported proved to need additional considerations as the expected precision can only be achieved in an very stable environment. The light seal of the enclosing unit has shown that a tighter fit would be beneficial. The calibration of the science images was found to collect undesired light from the environment over long exposures. 

The software written for this purpose, the Wavelength Scale Model, proved to have increased in complexity beyond the original plan. This was not unexpected but certainly challenging. Nonetheless, the spectral features could be identified to a fraction of a pixel in some cases and the computer code is ready to be adapted for further developments.  

The calibration of the spectrograph yielded a spectral resolution of R$\approx$50000, or expressed in radial velocity,

\begin{equation}
	RV=\frac{c}{R}\approx2\text{km}\text{s}^{-1}.
\end{equation} 

This means that to achieve a detection accuracy of $\sim$30ms$^{-1}$, which is the radial velocity expected from a Jupiter sized planet around a Sun-like star at the Earth's distance, we need to be able to measure a shift of the order of a 70$^{th}$ of a pixel.

The results obtained have laid the first steps of what is to be a long path to its full scale development. A stellar spectrum was captured and the most prominent solar absorption features detected to a sub-pixel accuracy.

\section{Future Work}
Several points have become clear after the tests performed during this honours project. At the instrumental level, an improved design in the light tight cage and thermal insulator could be addressed. This could lead to a more stable environment to acquire data, and an easier assembly/disassembly process.

The software in charge of the calibration of the spectrograph is currently using the coordinates of the main emission lines of mercury as a reference. The process of finding the list of coordinates is done manually by identifying the centroid of a given peak, and assigning the corresponding wavelength as a ``best guess''. Automating this process could streamline and add accuracy to the fitting process.

A robust image calibration pipeline needs to be implemented. This is currently being developed by Dr. Joao Bento and will largely increase the quality of images produced, leading to great benefits in later stages of the process. If several observing sites are to be considered, the automation of the observing site will need to be addressed. One of the key pillars of the replicable model is the capacity to acquire data with minimum human interaction. The number of candidate sites that are capable of participating in the project decays with the number of features that are required from them. This makes the expertise in automating a site a valuable asset to be able to deploy new spectrographs to a broader community. 

If this project is leading towards a large scale deployment, an organizational structure needs to be developed to support its several areas. One of the challenges will be to keep project coordination to a minimum to keep the budget focussed on science instead of administration. That can be achieved with efficient modularized components that can be mass produced and replaceable.

	\thispagestyle{empty}

\appendix
	\chapter{The Radial Velocity Equation}
\label{App:RV}

Initially, the two-body problem can be reduced to a single body by adjusting the semi-major axis using the reduce mass formula

\begin{equation}
	\displaystyle	a_1 = \left(\frac{m_2}{m_1+m_2}\right)a
\end{equation}  

where $a_1$ is the semi-major axis of the body analysed and $a=a_1+a_2$ is the maximum distance between the bodies. The distance from any point on the orbit to the centre of mass of the system can be written as 

\begin{equation}
	\displaystyle	r=\frac{a_1(1-e^2)}{1+e \cos f}
	\label{eq:r1}
\end{equation}  

where $e$ is the eccentricity and $f$ is the true anomaly\footnote{The angle formed by the position of the object, the centre of mass and the point in the orbit where the body is further from the centre of mass, or periapse.}, or 

\begin{equation}
	\displaystyle	r=\left(\frac{m_2}{m_1+m_2}\right)\frac{a(1-e^2)}{1+e\cos f}\textnormal{.}
\end{equation}

The value of $f$ cannot be computed analytically as a function of time and numerical solutions have to be used.

Adopting a Cartesian set of coordinates concentric to the barycentre with the $\unit{x}$-axis pointing in the direction of periastron, the position and velocity vectors are

\begin{equation}
	\mathbf{r}= \left(
		\begin{array}{c}
			\displaystyle	r\cos f  \\ [1ex]
			\displaystyle	r\sin f  
		\end{array} \right)
\end{equation},

and 

\begin{equation}
	\frac{d\mathbf{r}}{dt}= \left(
		\begin{array}{c}
				\displaystyle	\frac{dr}{dt}\cos f -r\frac{df}{dt}\sin f  \\ [2.5ex]
				\displaystyle	\frac{dr}{dt}\sin f + r\frac{df}{dt}\cos f  
		\end{array} \right)
		\label{eq:rdot}
\end{equation}
respectively.

Keeping in mind that the goal is to find the velocity as a function of $f$, $\frac{dr}{dt}$ and $\frac{df}{dt}$ need to be expressed as a function of $f$. 

Differentiating Equation~\ref{eq:r1} w.r.t. $t$ we obtain

\begin{equation}
	\displaystyle	\frac{dr}{dt}=\frac{a_1e(1-e^2)}{(1+e\cos f)^2}.
\end{equation} 

Simplifying with 

\begin{equation}
	\displaystyle	1+e \cos f=\frac{a_1(1-e^2)}{r}
\end{equation}  

from Equation~\ref{eq:r1}, we find

\begin{equation}
	\displaystyle	\frac{dr}{dt}=\frac{\displaystyle er^2\frac{df}{dt}\sin f}{a_1(1-e^2)}
\end{equation} 

and by replacing in Equation~\ref{eq:rdot} we obtain

\begin{equation}
	\displaystyle	\frac{d\mathbf{r}}{dt}= \left(
		\begin{array}{c}
				\displaystyle	\frac{e\,r^2\displaystyle\frac{df}{dt}\sin f}{a_1(1-e^2)}\cos f -r\frac{df}{dt}\sin f  \\ [2.5ex]
				\displaystyle	\frac{e\,r^2\displaystyle\frac{df}{dt}\sin f}{a_1(1-e^2)}\sin f + r\frac{df}{dt}\cos f 
		\end{array} \right)\textnormal{.}
	\label{eq:big_daddy}
\end{equation}

The following steps are a simplification of the Equation~\ref{eq:big_daddy}: 

\begin{equation}
	\frac{d\mathbf{r}}{dt}=r \frac{df}{dt} \left(
		\begin{array}{c}
				\displaystyle	\frac{e\,r\sin f}{a_1(1-e^2)}\cos f -\sin f  \\ [2.5ex]
				\displaystyle	\frac{e\,r\sin f}{a_1(1-e^2)}\sin f + \cos f 
		\end{array} \right)
\end{equation}
\vspace{5 mm}
\begin{equation}
	\frac{d\mathbf{r}}{dt}=r \frac{df}{dt} \left(
		\begin{array}{c}
				\displaystyle	\frac{e\,r\sin f}{a_1(1-e^2)}\cos f -\frac{r(1+e\cos f)\sin f}{a_1(1-e^2)}  \\ [2.5ex]
				\displaystyle	\frac{e\,r\sin f}{a_1(1-e^2)}\sin f + \frac{r(\cos f+e\cos^2 f)}{a_1(1-e^2)}
		\end{array} \right)
\end{equation}
\vspace{5 mm}
\begin{equation}
	\displaystyle\frac{d\mathbf{r}}{dt}=\frac{r^2\displaystyle \frac{df}{dt}}{a_1(1-e^2)} \left(
		\begin{array}{c}
				\displaystyle	e\sin f\cos f -(1+e\cos f)\sin f  \\ [2.5ex]
				\displaystyle	e\sin f\sin f + \cos f+e\cos^2 f
		\end{array} \right)
\end{equation}
\vspace{5 mm}
\begin{equation}
	\displaystyle\frac{d\mathbf{r}}{dt}=\displaystyle\frac{r^2\displaystyle\frac{df}{dt}}{a_1(1-e^2)}   \left(
		\begin{array}{c}
				\displaystyle	-\sin f  \\[1ex]
				\displaystyle	\cos f+e 
		\end{array} \right)\textnormal{.}
		\label{eq:rdot2}
\end{equation} 
\vspace{5 mm}

Energy and angular momentum are constants of motion of the system. Using $h_1=m_1r^2\frac{df}{dt}$ as the angular momentum, Equation~\ref{eq:rdot2} can be rewritten as 

\begin{equation}
	\frac{d\mathbf{r}}{dt}=\frac{h_1}{m_1a_1(1-e^2)}   \left(
		\begin{array}{c}
				\displaystyle	-\sin f  \\ [1ex]
				\displaystyle	\cos f+e 
		\end{array} \right)
	\label{eq:rdot3}
\end{equation}.

Using the expression

\begin{equation}
	\displaystyle	h=\displaystyle\sqrt{G(m_1+m_2)a(1-e^2)}.
\end{equation}

as the angular momentum of the system, $h_1$ can then be expressed in terms of $h$ by using the reduced mass:
\begin{equation}
	\displaystyle	h_1=\displaystyle\left(\frac{m_2}{m_1+m_2}\right)h=\displaystyle \sqrt{\displaystyle\frac{Gm_1^2m_2^4a(1-e^2)}{(m_1+m_2)^3}}.
\end{equation}

Replacing back in Equation~\ref{eq:rdot3} we find the general expression for the velocity as a function of $f$,

\begin{equation}
	\frac{d\mathbf{r}}{dt}=\sqrt{\frac{Gm_2^2}{(m_1+m_2)a(1-e^2)}}\left(
		\begin{array}{c}
				\displaystyle	-\sin f  \\ [1ex]
				\displaystyle	\cos f+e 
		\end{array} \right)
\end{equation}
\vspace{5 mm}

This equation is expressed in the frame of reference centred in the centre of mass of the system with the $\unit{x}$-axis pointing in the direction of periastron. To transform it into an equation that can be used to interpret observations from Earth, we need to find the projection of the velocity vector into the line of sight. 

The vector $\mathbf{k}$ can be described in terms of the frame of reference of the system, with the reminder that the $\unit{z}$ is perpendicular to the plane of the orbit and conforms to a right-hand convention. In such reference frame the $\mathbf{k}$ vector can be expressed as

\begin{equation}
	\mathbf{k}=\left(
		\begin{array}{c}
				\displaystyle	\sin\phi\sin \theta  \\ [1ex]
				\displaystyle	\cos\phi\sin \theta \\ [1ex]
				\displaystyle	\cos \theta
		\end{array} \right)
\end{equation}
\vspace{5 mm}
where $\theta$ and $\phi$ are the polar and azimuthal angles respectively. Then

\begin{equation}
	\begin{array}{rcl}
		\displaystyle	\frac{d\mathbf{r}}{dt}\cdot\mathbf{k}&=&\displaystyle \sqrt{\frac{G}{(m_1+m_2)a(1-e^2)}}m_2\sin \theta (\sin f\sin\phi+\cos f\cos\phi+e\cos\phi)\\ [3.5ex]
		\vspace{5 mm}
		\displaystyle	&=&\displaystyle \sqrt{\frac{G}{(m_1+m_2)a(1-e^2)}}m_2\sin \theta (\cos(\phi+f)+e\cos\phi)
	\end{array}
\end{equation}

We are interested in the radial velocity semi-amplitude,

\begin{equation}
	RV=
		\left(
			(\frac{d\mathbf{r}}{dt}
			\cdot
			\mathbf{k})_{max}
			-
			(\frac{d\mathbf{r}}{dt}
			\cdot
			\mathbf{k})_{min}
			\right)/2
\end{equation}
to finally yeild the radial velocity equation:
\vspace{5 mm}
\begin{equation}
	\displaystyle	RV=\sqrt{\frac{G}{(m_1+m_2)a(1-e^2)}}m_2\sin \theta  \textnormal{.}
\end{equation} 
	\chapter{Wavelength Scale Model}

\begin{lstlisting}
#Imports
import matplotlib.pyplot as plt     #python/matlab
import pylab
import random                       #random generator package
import pyfits
import os
import numpy as np
import matplotlib.cm as cm
import bisect as bis
import matplotlib.image as mpimg
import random

#least square package
from scipy.optimize.minpack import leastsq
from scipy import interpolate
from math import cos, sin, acos, asin, pi, atan, degrees, sqrt

#Astro Libraries
from astLib import astSED           

minLambda=0.5886 #min wavelength
maxLambda=0.59    #max wavelength
deltaLambda=0.0001    #step interval
maxLambda+=deltaLambda

#Can plot orders from  146 to 73 (about 390 to 795nm)
minOrder=146
maxOrder=73
deltaOrder=-1
maxOrder+=deltaOrder
booLog=6 
pixelSize= 5.4 

def main_errors(p, mainArgs):

    x,y,waveList,xSig,ySig = readCalibrationData(mainArgs[2])

    hdulist = pyfits.open('../c_noFlat_sky_0deg_460_median.fits')
    imWidth = hdulist[0].header['NAXIS1']
    imHeight = hdulist[0].header['NAXIS2']
    
    x=x-imWidth/2
    y=y-imHeight/2
    
    x_model, y_model, Lambda = main(p, mainArgs)
    
    x_best = x.copy()
    y_best = y.copy()
    for k in range(0,len(waveList)):
        ix, = np.where(waveList[k] == Lambda)
        if (ix.size == 0):
         x_best[k]=0
         y_best[k]=0
        else:
         best = ix[np.argmin(np.abs(y_model[ix] - y[k]))]
         x_best[k] = x_model[best]
         y_best[k] = y_model[best]

    return np.hstack([(x_best-x)/xSig,(y_best - y)/ySig]), waveList



def main( p, args):
    '''
    Compute the projection of n beams of monochromatic light 
    passing through an optical system. 

    Parameters
    ----------
    p : np np.array
        (beam phi, beam theta, prism1 phi, prism1 theta, 
        prism2 phi, prism2 theta, grating phi, grating theta,
         grating alpha, blaze period (microns), focal length(mm),
         distortion term) <-- optical arrangement
    args : np np.array
           (SEDMode(0=Max, 1=Random, 2=Sun, 3=from specFile, 
           			4=from CalibFile), 
           Plot?, specFile, Normalize intensity? (0=no, #=range),
           Distort?, Interpolate, PlotCalibPoints) <-- other options
   
    Returns
    -------
    x : np np.array
        x coordinate of the target point 
    y : np np.array
        x coordinate of the target point 
    lambda : np np.array
        wavelength at x,y
 
    '''  
    
    global n1, n2, n4, n5, s, l, d, flux, 
    global booPlot, specFile, booPlotCalibPoints, booInterpolate,
    global booGaussianFit plotBackImage
    global allFlux,booPlotLabels, specFile
    
    #Args breakdown
    SEDMode = int(args[0])
    booPlot = int(args[1])
    specFile = args[2]
    intNormalize = int(args[3])
    booDistort = int(args[4])
    booInterpolate=int(args[5])
    booPlotCalibPoints=int(args[6])
    booPlotLabels=int(args[7])
    plotBackImage=args[8]
    booGaussianFit=int(args[9])
    
    #Initial beam
    uiphi = p[0]*pi/180    #'Longitude' with the x axis as 
    uitheta = p[1]*pi/180  #Latitude with the y axis the polar axis
    u=np.array([cos(uiphi)*sin(uitheta),
    				sin(uiphi)*sin(uitheta),
    				cos(uitheta)])
       
    #Focal length
    fLength = p[10]
    
    #Prism surface 1
    n1phi = p[2]*pi/180   
    n1theta = p[3]*pi/180 
    n1=np.array([cos(n1phi)*sin(n1theta),
    				 sin(n1phi)*sin(n1theta),
    				 cos(n1theta)])
    
    #Prism surface 2
    n2phi = p[4]*pi/180   
    n2theta = p[5]*pi/180 
    n2=np.array([cos(n2phi)*sin(n2theta),
    				 sin(n2phi)*sin(n2theta),
    				 cos(n2theta)])

    #Prism surface 3 (surf #2 on return)
    n4=-n2
    
    #Prism surface 4 (surf #1 on return)
    n5=-n1 
    
    #Grating
    d = p[9]  #blaze period in microns  
    sphi = p[6]*pi/180   
    stheta = p[7]*pi/180 
    s = np.array([cos(sphi)*sin(stheta),
    				  sin(sphi)*sin(stheta),
    				  cos(stheta)]) #component perp to grooves
        
    #Now find two vectors perpendicular to s:
    a = np.array([s[1]/np.sqrt(s[0]**2 + s[1]**2), 
    				 -s[0]/np.sqrt(s[0]**2 + s[1]**2), 0])
    b = np.cross(a,s)
    
    #Create l from given alpha using a and b as basis
    alpha = p[8]*pi/180 
    l = cos(alpha)*a + sin(alpha)*b #component along grooves
       
    #Distortion np.array
    K = p[11]
       
    #Launch grid loop. Creates an array of (x,y,lambda)
    CCDMap = doCCDMap(u ,minLambda ,maxLambda ,deltaLambda ,
				minOrder ,maxOrder ,deltaOrder ,fLength,
				stheta, SEDMode, intNormalize) 
        
    #Distort
    if booDistort==1:
        x=CCDMap[:,0]
        y=CCDMap[:,1]
        CCDMap[:,0], CCDMap[:,1] = distort(x, y, K)
    
    #Validates output
    x=y=Lambda=0
    if CCDMap.size>0:
        x=CCDMap[:,0]
        y=CCDMap[:,1]
        Lambda=CCDMap[:,2]
        
    #Plot
    if booPlot==1:
        doPlot(CCDMap)
    
    return x, y, Lambda


def extractOrder(x,y,image):
    
    flux=np.zeros(len(y))
    flux2=np.zeros(len(y))
    flux3=np.zeros(len(y))

    hdulist = pyfits.open(image)
    imWidth = hdulist[0].header['NAXIS1']
    imHeight = hdulist[0].header['NAXIS2']
    
    im = pyfits.getdata(image)  

    x=x+imWidth/2
    y=y+imHeight/2

    for k in range(0,len(y)):
        x_int = round(x[k])
        in_image_temp = im[y[k],x_int-5:x_int+6]
        in_image_temp[in_image_temp < 0] = 0
        xv = np.arange(-5,6) - x[k] + x_int
        flux[k] =  np.sum(in_image_temp * np.exp(-(xv/3.5)**4))

    return flux, flux2 +np.average(flux), flux3


def fftshift1D(inImage, shift):
    '''
    This program shifts an image by sub-pixel amounts.
       
    Parameters
    ----------
    inImage :  image
        Input image
        
    shift : array
        (x,y) pixel shift
        
    Returns
    -------
    outImage : Image
        Shifted Image
    '''  
    
    ftin = np.fft.fft(inImage)
    sh = inImage.shape
    
    #The following line makes a meshgrid np.array as floats.
    xy = np.mgrid[0:sh[0]] + 0.0
    xy[:] = (((xy[0,:] + sh[0]/2) % sh[0]) - sh[0]/2)/float(sh[0])

	db = np.real(
		np.fft.ifft(
			ftin*np.exp(
				np.complex(0,-2*np.pi)
				*(xy[0,:,:]*shift[0] + 				
				xy[1,:,:]*shift[1]))))
	return db
    
    
def doCCDMap(u, minLambda, maxLambda, deltaLambda, 
			minOrder, maxOrder, deltaOrder, fLength, 
			stheta, SEDMode, intNormalize):
    
    dataOut=np.zeros((1,5))
    
    #Loads SEDMap based on selection. 
    SEDMap = doSEDMap(SEDMode, minLambda, maxLambda, 
    						deltaLambda, intNormalize)
    						
    blaze_angle = stheta #Approximately atan(2)
    allFlux = np.array([0])
    allLambdas = np.array([0])
    
    '''Main loop
    Navigates orders within the range given
    For each order navigates the list of wavelenghts in SEDMap
    '''    
    for nOrder in range(minOrder, maxOrder, deltaOrder):

        LambdaBlMin = 2*d*sin(blaze_angle)/(nOrder+1)
        LambdaBlMax = 2*d*sin(blaze_angle)/(nOrder-1)

        SEDMapLoop=SEDMap.copy()
        
        #constrain by +/- FSP (was FSP/2)
        SEDMapLoop = SEDMapLoop[SEDMapLoop[:,0]>=LambdaBlMin]
        if SEDMapLoop.shape[0]>0:       
            SEDMapLoop = SEDMapLoop[SEDMapLoop[:,0]<=LambdaBlMax]     

        #loop lambda for current order
        for Lambda,inI in SEDMapLoop: 
                                    
            nPrism = nkzfs8(Lambda)
            nAir = n(Lambda)
            
            #Computes the unit vector that results 
             from the optical system for a given wavelength 
             and order
            v, isValid = rayTrace(nAir, nPrism, nOrder, Lambda, 
            			d, u, n1, n2, n4, n5, s, l)
            
            if isValid: #no errors in calculation
            		# coordinates in focal plane in pixels
                x=v[0]*fLength*1000/pixelSize 
                z=v[2]*fLength*1000/pixelSize 
    
                outI=Intensity(Lambda, minLambda, maxLambda)              
                dataOut= np.vstack(
                		(dataOut,np.array(
                			[x,z, Lambda, inI*outI ,nOrder]))) 
        
        #Order extraction        
        if (booInterpolate==1 and 
        		len(dataOut[dataOut[:,4]==nOrder][:,0])>=3):
            
            xPlot=dataOut[dataOut[:,4]==nOrder][:,0]
            yPlot=dataOut[dataOut[:,4]==nOrder][:,1]
            LambdaPlot=dataOut[dataOut[:,4]==nOrder][:,2]
            
            fLambda = interpolate.interp1d(yPlot, LambdaPlot)
            fX = interpolate.interp1d(yPlot, xPlot, 
            			'quadratic', bounds_error=False)
          
            hdulist = pyfits.open(plotBackImage)
            imWidth = hdulist[0].header['NAXIS1']
            imHeight = hdulist[0].header['NAXIS2']
            
            newY = np.arange(-imHeight/2,imHeight/2)
            
            newX = fX(newY)
            
            nanMap = np.isnan(newX)
            newX = newX[-nanMap]
            newY = newY[-nanMap]
                    
            flux,flux2,flux3 = extractOrder(newX,newY,plotBackImage)
            
            #read flats
            image = '../simple_flat.fits'
            fluxFlat,flux2,flux3 = extractOrder(newX,newY,image)

            Lambdas = fLambda(newY)
            
            #Blackbody curve to balance flats
            BB = Lambdas**(-4) / (np.exp(14400/Lambdas/3000)- 1)

            cleanFlux = flux/fluxFlat*BB#/nOrder**2

            # Fit a Gaussian             
            if booGaussianFit==1: 
            
                X=Lambdas.copy()
                Y=cleanFlux.copy()
                if len(Y)>0:
            			a,FWHMIndex = find_nearest(Y,np.max(Y)/2)
                    maxIndex =n p.where(Y==np.max(Y))[0][0]
                    FWHM = 2*(X[maxIndex]-X[FWHMIndex])
                    fit_mu = X[maxIndex]
                    R=fit_mu/FWHM

                    plt.plot(X,Y) 
                    plt.ylabel('Intensity (Relative Units)')
                    plt.xlabel('Wavelength (Micrometers)')
                    plt.title('Spectral Resolution at '+
                    		 str("{:0.4f}".format(fit_mu))+'
                    		  micrometers')
                    plt.annotate('FWHM='+
                    		str("{:0.4f}".format(FWHM*1000))+
                    		'nm R='+
                    		str("{:0.4f}".format(fit_mu/FWHM)),
                    		xy = (X[0],Y[0]), 
                    		xytext = (220, 250),
                    		textcoords = 'offset points', 
                    		ha = 'right', 
                    		va = 'bottom',
                    		bbox = dict(
                    			 boxstyle = 'round,pad=0.5',
                    			  fc = 'white', 
                    			  alpha = 0.9), 
                    			  size=15)
                    plt.axvspan(fit_mu-FWHM/2, 
                    		fit_mu+FWHM/2, 
                    		facecolor='g', 
                    		alpha=0.5)
                    plt.show()

            
            if np.sum(allFlux)>0:
                intersectStart=bis.bisect(allLambdas,np.min(Lambdas))   
                intersectEnd=len(allLambdas)   
                bestDistance=1e10     
                bestIndex=0
                
                for k in range(0,intersectEnd-intersectStart):
                    currDistance=sqrt((allFlux[intersectStart+k]
                    				- cleanFlux[k])**2)
                    if currDistance<bestDistance:
                        bestDistance = currDistance
                        bestIndex = k
                        
                allLambdas=allLambdas[allLambdas<Lambdas[bestIndex]]
                allFlux=allFlux[allLambdas<Lambdas[bestIndex]]            
                allLambdas=np.hstack((allLambdas,Lambdas[bestIndex:]))
                allFlux=np.hstack((allFlux,cleanFlux[bestIndex:]))
                
            else:
                allLambdas=Lambdas
                allFlux=cleanFlux
    

    if booInterpolate==1:   
        fig = plt.figure()
        ax1 = fig.add_subplot(111)
        ax1.plot(allLambdas,allFlux)
        plt.title('Sodium Doublet')
        plt.ylabel('Intensity (Relative Units)')
        plt.xlabel('Wavelength (Micrometers)')
        plt.show() 
       
    CCDMap=dataOut[1:,]

    return CCDMap


def gauss(x, p): # p[0]==mean, p[1]==stdev
	result = 1.0/(p[1]*np.sqrt(2*np.pi))*
				np.exp(-(x-p[0])**2/(2*p[1]**2))
    return result

def doPlot(CCDMap):
        x = CCDMap[:,0] 
        z = CCDMap[:,1] 
        Lambda = CCDMap[:,2] 
        Intensity= CCDMap[:,3] 

        colorTable = np.array((wav2RGB(Lambda, Intensity))) 
        
        hdulist = pyfits.open(plotBackImage)
        imWidth = hdulist[0].header['NAXIS1']
        imHeight = hdulist[0].header['NAXIS2']

        im = pyfits.getdata(plotBackImage)
        im[im<0]=0
        im /= im.max()
        im = np.sqrt(im) #Remove this line for Hg

        fig = plt.figure()
        ax1 = fig.add_subplot(111)

        plt.imshow(im,extent=[-imWidth/2, imWidth/2, 
        				-imHeight/2, imHeight/2])
        plt.set_cmap(cm.Greys_r)
        ax1.scatter(x, -z ,s=8, 
        				color=colorTable , 
        				marker='o', alpha =.5)

        if booPlotLabels==1:
            for label, x, y in zip(Lambda, x, -z):
                plt.annotate(
                    label, 
                    xy = (x, y), 
                    xytext = (0,-20),
                    textcoords = 'offset points', 
                    ha = 'right', 
                    va = 'bottom',
                    bbox = dict(
                    		   boxstyle = 'round,pad=0.5', 
                    		   fc = 'white', 
                    		   alpha = 0.9),
                    arrowprops = dict(
                    			arrowstyle="wedge,tail_width=1.",
                    			fc=(0, 0, 1), 
                    			ec=(1., 1, 1),
                    			patchA=None,
                    			relpos=(0.2, 0.8),                   
                    			connectionstyle="arc3,rad=-0.1"),
                    			size=7)

        plt.ylabel('pixels')
        plt.xlabel('pixels')
        
        
        
        if booPlotCalibPoints==1:
            x,y,waveList,xSig,ySig = readCalibrationData(specFile)
            ax1.scatter(x-imWidth/2 ,
            			    -(y-imHeight/2) ,
            			    s=400, color='black', 
            			    marker='x', alpha=1)

        plt.title('Order Identification')

        plt.axis([-imWidth/2 , imWidth/2 , 
        			  -imHeight/2 , imHeight/2])
        
        plt.show()


def rayTrace(nAir, nPrism, nOrder, Lambda, 
				d, u, n1, n2, n4, n5, s, l):
    ''' 
    Traces a beam through the spectrograph. 
    Spectrograph frame of reference.
    From the opposite end of the camera looking at the camera
    x=to the right, y=to camera, z=up
    u*=beam, n*=surface normals
    s=grating, perp to the grooves.
    l=grating, parallel to the grooves.
    d=blaze period
    '''
                          
    #Vector transform due to first surface
    u = Snell3D(nAir, nPrism, u, n1)
 
    #Vector transform due to second surface
    u = Snell3D(nPrism, nAir, u, n2)
    
    #Vector transform due to grating             
    u, isValid = Grating(u, l, s, nOrder, Lambda, d)
         
    if isValid:
        #Vector transform due to third surface
        u = Snell3D(nAir, nPrism, u, n4)
        
        #Vector transform due to fourth surface
        u = Snell3D(nPrism, nAir, u, n5)
         

    return u, isValid


def Snell3D(n_i,n_r,u,n):
    '''
    Computes the new direction of a vector when changing medium. 
    n_i, n_r = incident and refractive indices
    '''
 
    u_p = u - np.dot(u,n)*n
    u_p /= np.linalg.norm(u_p)
    
    theta_i = acos(np.dot(u,n))
    
    if n_i*sin(theta_i)/n_r<=1:
        theta_f = asin(n_i*sin(theta_i)/n_r)
        u = u_p*sin(pi-theta_f) + n*cos(pi-theta_f)    

    return u       


def Grating(u, l, s, nOrder, Lambda, d):
    #Computes the new direction of a vector when hitting a grating.
    
    isValid=False

    n = np.cross(s, l) 
    u_l = np.dot(u, l)
    u_s = np.dot(u, s) + nOrder*Lambda/d  
    if (1-u_l**2 -u_s**2)>=0: 
        u_n = np.sqrt(1- u_l**2 - u_s**2)
        u = u_l*l + u_s*s + u_n*n
        isValid=True

    return u, isValid     


def doSEDMap(SEDMode, minLambda, maxLambda, deltaLambda, intNormalize):
    '''
    Loads the SED map
       
    Parameters
    ----------
    SEDMode : int
        Mode for the creation of the SEDMap
        0=Max, 1=Random, 2=Sun, 
        3=from specFile, 4=from Calibration file
        
    minLambda : np.float32
        Lower limit for SEDMap.

    maxLambda : np.float32
        Higher limit for SEDMap.

    deltaLambda : np.float32
        Step between wavelengths.
    
    intNormalize : integer
        if !=0, it normalizes to intNormalize value
        
    Returns
    -------
    SEDMap : np np.array
        n x 2 np np.array with wavelength, Energy
 
    '''  
    if SEDMode==0: #Flat
        SEDMap = np.column_stack((np.arange(minLambda, 
        				maxLambda, 
        				deltaLambda),
        				np.ones(np.arange(minLambda,
        						maxLambda,
        						deltaLambda).size)))

    elif SEDMode==1: #Random
        SEDMap = np.array([0,0])
        Lambdas = range(minLambda, maxLambda, deltaLambda)
        
        for Lambda in Lambdas:
            SEDMap = np.vstack((
            		SEDMap,np.array([Lambda,
            				random.random(0.0,1.0)])))
            
        SEDMap = SEDMap[1:,]
                 
    elif SEDMode==2: #Sun
        sol = astSED.SOL               
        tempA=sol.wavelength.transpose()*1e-4
        tempB=sol.flux.transpose()            
        SEDMap = np.column_stack((tempA, tempB))    
        
        #Remove rows outside the wavelength range
        SEDMap = SEDMap[SEDMap[:,0]>=minLambda]     
        SEDMap = SEDMap[SEDMap[:,0]<=maxLambda]     
                                  
    elif SEDMode==3: #From file
        SEDMap = np.array([0,0])
         
        for line in open(specFile):
            Lambda = float(str(line).split()[0]) #Wavelength
            I = float(str(line).split()[1])       #Intensity
            SEDMap = np.vstack((SEDMap,np.array([Lambda,I])))

        SEDMap=SEDMap[1:,]  
         
    elif SEDMode==4: #From calibration file
        SEDMap = np.array([0,0])
         
        for line in open(specFile):
            Lambda = float(str(line).split()[2]) #Wavelength
            I = 1      #Intensity
            SEDMap = np.vstack((SEDMap,np.array([Lambda,I])))    
        
        SEDMap=SEDMap[1:,]  
                
    return SEDMap


def wav2RGB(Lambda, Intensity):
    '''Converts Lambda into RGB'''
    
    out=np.array([0,0,0])
    
    for i in range(Lambda.size):
    
        w = Lambda[i]
        I = Intensity[i]
    
        # colour
        if w >= .380 and w < .440:
            R = -(w - .440) / (.440 - .350)
            G = 0.0
            B = 1.0
        elif w >= .440 and w < .490:
            R = 0.0
            G = (w - .440) / (.490 - .440)
            B = 1.0
        elif w >= .490 and w < .510:
            R = 0.0
            G = 1.0
            B = -(w - .510) / (.510 - .490)
        elif w >= .510 and w < .580:
            R = (w - .510) / (.580 - .510)
            G = 1.0
            B = 0.0
        elif w >= .580 and w < .645:
            R = 1.0
            G = -(w - .645) / (.645 - .580)
            B = 0.0
        elif w >= .645 and w <= .780:
            R = 1.0
            G = 0.0
            B = 0.0
        else:
            R = 1.0
            G = 1.0
            B = 1.0
    
        # intensity correction
        if w >= .3800 and w < .4200:
            SSS = 0.3 + 0.7*(w - .3500) / (.4200 - .3500)
        elif w >= .4200 and w <= .7000:
            SSS = 1.0
        elif w > .7000 and w <= .7800:
            SSS = 0.3 + 0.7*(.7800 - w) / (.7800 - .7000)
        else:
            SSS = 1.0
        SSS *= (I)

        out=np.vstack((out,np.array( 
        		[float(SSS*R), float(SSS*G), float(SSS*B)]))) 
        
    return out[1:,]


def Intensity(Lambda, minLambda, maxLambda):
    '''
    Retrieves or calculates the expected relative intensity 
    based on distance from the central lambda value
       
    Parameters
    ----------
    Lambda : np.float32
        Wavelength.
    minLambda : np.float32
        Lower end of wavelength range.
    maxLambda : np.float32
        Higher end of wavelength range. 
   
    Returns
    -------
    z : np.float32
        result 0 to 1.
    '''

    x = (((float(Lambda) - float(minLambda))
    		/(float(maxLambda) - float(minLambda)))-0.5)*2
    if x!=0:
        z=sin(x*pi)/(x*pi)

    return z


def n(Lambda, t=18, p=101325):

    n = (0.0472326 * (173.3 - (1/Lambda)**2)**(-1))+1
    
    return n


def findFit(calibrationFile, p_try, factor_try ,diag_try):
    '''
    Wrapper for reading the calibration file, 
    and launching the fitting function
       
    Parameters
    ----------
    calibrationFile : string
        Name of the file with the data from the spectrograph

    Returns
    -------
    fit : np np.array
        1 x 12 np np.array with fitted arguments (p np.array)
    '''  
    
    mainArgs = ['4','0',calibrationFile,
    				'0','0','0','0','1',
    				'../c_noFlat_sky_0deg_460_median.fits']   
    
    fit = leastsq(main_errors,p_try, 
    				  args=mainArgs, 
    				  full_output=True, 
    				  factor=factor_try, 
    				  diag=diag_try)

    return fit
   

\end{lstlisting}
	\chapter{Grating Orientation}
\label{App:gr_orientation}

\begin{figure}[H]
	\begin{center}
		\setlength\fboxsep{0pt}
		\setlength\fboxrule{0pt}
		\fbox{\includegraphics[width=300px]{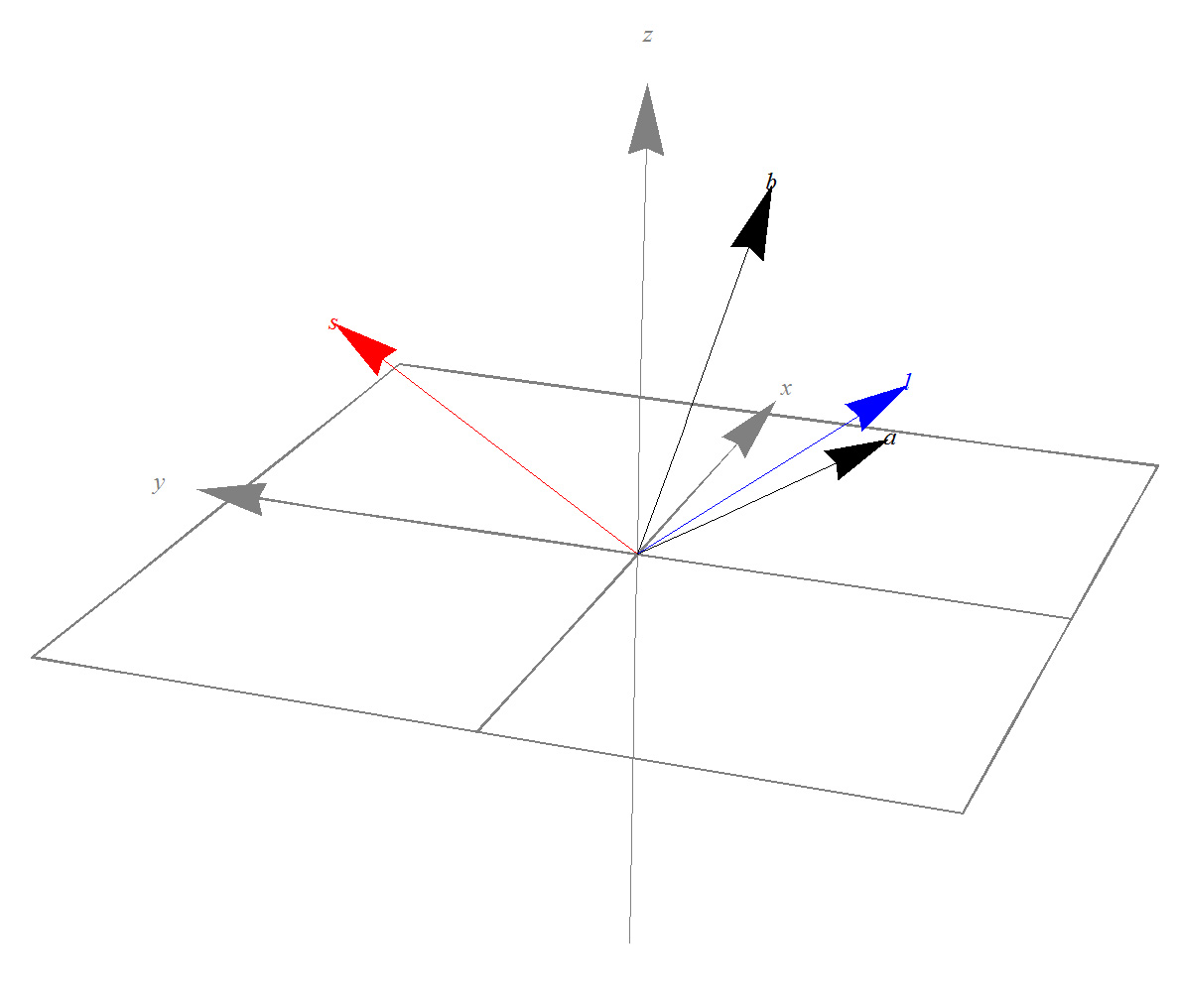}}
	\end{center}
	\caption{Vector $\unit{l}$ as a function of $\unit{a}$ and $\unit{b}$}
	\label{fig:aandb}
\end{figure}

The unique orientation of the grating requires 3 angles to be identified. The steps to construct the 2 vectors that achieve this are described here. 

\subsubsection{The s vector}

From the provided polar and azimuthal angles, $\phi$ and $\theta$, the $\unit{s}$ vector can be constructed by a simple coordinate transformation

\begin{equation}
	\unit{s}=(\cos \phi \sin\theta , \sin\phi \sin\theta , \cos\theta)
	\nonumber
	\label{eq:s}
\end{equation} 

\subsubsection{The l vector}

To find the $\unit{l}$ vector 2 steps are needed. First we need to find a set of basis that span the plane perpendicular to $\unit{s}$. Second, define the orientation of $\unit{l}$ as a linear combination of these basis. 

The vectors $\unit{a}$ and $\unit{b}$ are introduced as auxiliary vectors. To span the plane perpendicular to $\unit{s}$ they need to be perpendicular to each other. The derivation is found in Appendix \ref{App:gr_orientation}. 

As an initial constraint, the vector $\unit{a}$ is defined to live in the x-y plane (i.e. $a_z=0$). With this constrain, a unique vector can be found that satisfies the following conditions:

\begin{equation}
	\unit{a} \perp \unit{s},
	\nonumber
\end{equation} 

\begin{equation}
	\unit{a}\perp \unit{z},
	\nonumber
\end{equation} 

\begin{equation}
	|\unit{a}|=1.
	\nonumber
\end{equation} 

We know that

\begin{equation}
	\boxed{a_z=0},
	\nonumber
\end{equation} 

so

\begin{equation}
	a_x s_x+a_y s_y=0
	\label{eq:adots}
\end{equation} 

\begin{equation}
	a_x^2+a_y^2=1
	\label{eq:moda}
\end{equation} 

rearranging \ref{eq:adots} and \ref{eq:moda},

\begin{equation}
	a_x s_x=-a_y s_y
	\label{eq:adots2}
\end{equation} 

\begin{equation}
	a_y=\sqrt{1-a_x^2},
	\label{eq:moda2}
\end{equation} 

combining \ref{eq:adots2} and \ref{eq:moda2} and rearranging for $a_x$:

\begin{equation}
	a_x s_x=\sqrt{1-a_x^2} s_y
	\nonumber
\end{equation} 

\begin{equation}
	a_x^2 s_x^2=(1-a_x^2) s_y^2
	\nonumber
\end{equation} 

\begin{equation}
	a_x^2 (s_x^2+s_y^2)=s_y^2
	\nonumber
\end{equation} 

\begin{equation}
	\boxed{a_x=\frac{s_y}{\sqrt{(s_x^2+s_y^2)}}}
	\nonumber
\end{equation} 

replacing \ref{eq:adots}

\begin{equation}
	\frac{s_y}{\sqrt{(s_x^2+s_y^2)} s_x+a_y s_y}=0
	\nonumber
\end{equation}

\begin{equation}
	\boxed{a_y=-\frac{s_x}{\sqrt{(s_x^2+s_y^2)}}}
	\nonumber
\end{equation} 

so we find explicit from of $\unit{a}$ as a function of $\unit{s}$:

\begin{equation}
	\boxed{\unit{a}=(\frac{s_y}{\sqrt{(s_x^2+s_y^2)}} ,-\frac{s_x}{\sqrt{(s_x^2+s_y^2)}}, 0)}
	\nonumber
\end{equation} 

The vector $\unit{b}$ is simply the cross product between $\unit{a}$ and $\unit{s}$.

\begin{equation}
	\boxed{\unit{b}=\unit{a} \times \unit{s}}
	\nonumber
\end{equation} 

Having defined the basis to describe the $\unit{l}$ vector we find:
 
\begin{equation}
	\boxed{\unit{l}=\cos\alpha \unit{a} + \sin\alpha \unit{b}}
	\nonumber
\end{equation} 

where the angle $\alpha$ is one of the parameters and it is measured from $\unit{a}$ to $\unit{b}$.


\bibliography{Compact_Spectrograph}{}
\end{document}